\documentclass[12pt]{article}
\usepackage{graphics,graphicx}
\usepackage{amssymb,epsfig,amsmath,euscript,array}
\usepackage[nosort]{cite}
\usepackage{axodraw4j}
\usepackage{pstricks}
\usepackage{color}

\makeatletter
\@addtoreset{equation}{section}
\makeatother

\newcounter{multieqs}

\newcommand{\be}{\begin{equation}}
\newcommand{\ee}{\end{equation}}

\newcommand{\bm}[1]{\mbox{\boldmath $#1$}}

\newcommand{\kslash}{k \!\!\! / }

\newcommand{\lslash}{l \!\! / }
\newcommand{\Pslash}{P \!\!\!\! / }

\newcommand{\islash}{i \!\!\! / }
\newcommand{\jslash}{j \!\!\! / }
\newcommand{\aslash}{a \!\!\! / }
\newcommand{\bslash}{{b \hspace{-6pt} \slash} }

\newcommand{\onslash}{1 \!\!\! / }
\newcommand{\twslash}{2 \!\!\!/ }
\newcommand{\thslash}{3 \!\!\!/ }
\newcommand{\foslash}{4 \!\!\! / }
\newcommand{\fislash}{5 \!\!\! / }

\newcommand{\mslash}{m \!\!\! / }

\def\bd{\begin{document}}
\def\ed{\end{document}}
\def\nn{\nonumber}
\def\bea{\begin{eqnarray}}
\def\eea{\end{eqnarray}}

\def\ab{(ijab)}
\def\ba{(ijba)}
\def\ijab{{\tr}_{+}(\islash\, \jslash\, \aslash \, \bslash)}
\def\ijba{{\tr}_{+}(\islash\, \jslash\, \bslash \, \aslash)}
\def\ijaP{{\tr}_{+}(\islash\, \jslash\, \aslash \, \Pslash)}
\def\ijPLa{{\tr}_{+}(\islash\, \jslash\, \Pslash_L \, \aslash)}
\def\ijaPL{{\tr}_{+}(\islash\, \jslash\, \aslash \, \Pslash_L)}
\def\ijPLza{{\tr}_{+}(\islash\, \jslash\, \Pslash_{L;z} \, \aslash)}
\def\ijaPLz{{\tr}_{+}(\islash\, \jslash\, \aslash \, \Pslash_{L;z})}
\def\ijPa{{\tr}_{+}(\islash\, \jslash\, \Pslash \, \aslash)}
\def\iaPb{{\tr}_{+}(\islash\, \aslash\, \Pslash \, \bslash)}
\def\ibPa{{\tr}_{+}(\islash\, \bslash\, \Pslash \, \aslash)}
\def\ijPmu{{\tr}_{+}(\islash\, \jslash\, \Pslash \, \mu)}
\def\ibmuP{{\tr}_{+}(\islash\, \bslash\, \mu \, \Pslash)}
\def\ibmua{{\tr}_{+}(\islash\, \bslash\, \mu \, \aslash)}
\def\iamub{{\tr}_{+}(\islash\, \aslash\, \mu \, \bslash)}
\def\jaPb{{\tr}_{+}(\jslash\, \aslash\, \Pslash \, \bslash)}
\def\ijmuP{{\tr}_{+}(\islash\, \jslash\, \mu \, \Pslash)}
\def\ijmum{{\tr}_{+}(\islash\, \jslash\, \mu \, \mslash)}
\def\ijmmu{{\tr}_{+}(\islash\, \jslash\, \mslash \, \mu)}
\def\ijmP{{\tr}_{+}(\islash\, \jslash\, \mslash \, \Pslash)}
\def\iabP{{\tr}_{+}(\islash\, \aslash\, \bslash \, \Pslash)}
\def\ijbP{{\tr}_{+}(\islash\, \jslash\, \bslash \, \Pslash)}
\def\jbPa{{\tr}_{+}(\jslash\, \bslash\, \Pslash \, \aslash)}
\def\ijPb{{\tr}_{+}(\islash\, \jslash\, \Pslash \, \bslash)}
\def\jbmua{{\tr}_{+}(\jslash\, \bslash\, \mu \, \aslash)}

\def\loablt{ {\tr}_{+}(\lslash_1\, \aslash \, \bslash\, \lslash_2)}

\def\ijlolt{{\tr}_{+}(\islash\, \jslash\, \lslash_1 \, \lslash_2)}
\def\ijltlo{{\tr}_{+}(\islash\, \jslash\, \lslash_2 \, \lslash_1)}
\def\ibloa{{\tr}_{+}(\islash\, \bslash\, \lslash_1 \, \aslash)}
\def\jaltb{{\tr}_{+}(\jslash\, \aslash\, \lslash_2 \, \bslash)}
\def\ialtb{{\tr}_{+}(\islash\, \aslash\, \lslash_2 \, \bslash)}
\def\bltloa{{\tr}_{+}(\bslash\, \lslash_2\, \lslash_1 \, \aslash)}
\def\jbloa{{\tr}_{+}(\jslash\, \bslash\, \lslash_1 \, \aslash)}
\def\ibPb{{\tr}_{+}(\islash\, \bslash\, \Pslash \, \bslash)}
\def\ijltb{{\tr}_{+}(\islash\, \jslash\, \lslash_2 \, \bslash)}

\def\ijloa{{\tr}_{+}(\islash\, \jslash\,  \lslash_1 \, \aslash)}
\def\ijblt{{\tr}_{+}(\islash\, \jslash\,  \bslash \, \lslash_2)}

\def\jakb{{\tr}_{+}(\jslash\, \aslash\, \kslash \, \bslash)}
\def\iakb{{\tr}_{+}(\islash\, \aslash\, \kslash \, \bslash)}

\def\tofo{{\tr}_{+}(\onslash\, \thslash\, \twslash \, \foslash)}
\def\foto{{\tr}_{+}(\onslash\, \thslash\, \foslash \, \twslash)}
\def\tofi{{\tr}_{+}(\onslash\, \thslash\, \twslash \, \fislash)}
\def\fito{{\tr}_{+}(\onslash\, \thslash\, \fislash \, \twslash)}

\def\lrangle#1#2{\langle #1\,#2\rangle}

\def\Li{{$\rm Li}_2$}
\def\eps{\epsilon}
\def\epsuv{{\epsilon_{\rm \mbox{\tiny UV}}}}
\let\bm=\bibitem
\let\la=\label

\def\npb#1#2#3{Nucl. Phys. {\bf{B#1}} #3 (#2)}
\def\plb#1#2#3{Phys. Lett. {\bf{#1B}} #3 (#2)}
\def\prl#1#2#3{Phys. Rev. Lett. {\bf{#1}} #3 (#2)}
\def\prd#1#2#3{Phys. Rev. {D \bf{#1}} #3 (#2)}
\def\cmp#1#2#3{Comm. Math. Phys. {\bf{#1}} #3 (#2)}
\def\cqg#1#2#3{Class. Quantum Grav. {\bf{#1}} #3 (#2)}
\def\nppsa#1#2#3{Nucl. Phys. B (Proc. Suppl.) {\bf{#1A}}#3 (#2)}
\def\ap#1#2#3{Ann. of Phys. {\bf{#1}} #3 (#2)}
\def\ijmp#1#2#3{Int. J. Mod. Phys. {\bf{A#1}} #3 (#2)}
\def\rmp#1#2#3{Rev. Mod. Phys. {\bf{#1}} #3 (#2)}
\def\mpla#1#2#3{Mod. Phys. Lett. {\bf A#1} #3 (#2)}
\def\jhep#1#2#3{J. High Energy Phys. {\bf #1} #3 (#2)}
\def\atmp#1#2#3{Adv. Theor. Math. Phys. {\bf #1} #3 (#2)}
\newcommand{\EQ}[1]{\begin{equation} #1 \end{equation}}
\newcommand{\AL}[1]{\begin{subequations}\begin{align} #1 \end{align}\end{subequations}}
\newcommand{\SP}[1]{\begin{equation}\begin{split} #1 \end{split}\end{equation}}
\newcommand{\ALAT}[2]{\begin{subequations}\begin{alignat}{#1} #2 \end{alignat}
                        \end{subequations}}
\def\beqa{\begin{eqnarray}}
\def\eeqa{\end{eqnarray}}
\def\beq{\begin{equation}}
\def\eeq{\end{equation}}
\def\sst{\scriptscriptstyle}
\def\thetabar{\bar\theta}
\def\Tr{{\rm Tr}}
\def\one{\mbox{1 \kern-.59em {\rm l}}}
 \def\Nh{\hat{N}}

\newcommand{\half}{{\textstyle {1 \over 2}}}

\def\a{\alpha}      \def\da{{\dot\alpha}}
\def\b{\beta}       \def\db{{\dot\beta}}
\def\c{\gamma}  \def\G{\Gamma}  \def\cdt{\dot\gamma}
\def\d{\delta}  \def\D{\Delta}  \def\ddt{\dot\delta}
\def\e{\epsilon}        \def\vare{\varepsilon}
\def\f{\phi}    \def\F{\Phi}    \def\vvf{\f}
\def\h{\eta}
\def\k{\kappa}
\def\l{\lambda} \def\L{\Lambda}
\def\m{\mu} \def\n{\nu}
\def\o{\omega}
\def\p{\pi} \def\P{\Pi}
\def\r{\rho}
\def\s{\sigma}  \def\S{\Sigma}
\def\t{\tau}
\def\th{\theta} \def\Th{\Theta} \def\vth{\vartheta}
\def\X{\Xeta}
\def\z{\zeta}
\def\de{\partial}

\def\cA{{\cal A}} \def\cB{{\cal B}} \def\cC{{\cal C}}
\def\cD{{\cal D}} \def\cE{{\cal E}} \def\cF{{\cal F}}
\def\cG{{\cal G}} \def\cH{{\cal H}} \def\cI{{\cal I}}
\def\cJ{{\cal J}} \def\cK{{\cal K}} \def\cL{{\cal L}}
\def\cM{{\cal M}} \def\cN{{\cal N}} \def\cO{{\cal O}}
\def\cP{{\cal P}} \def\cQ{{\cal Q}} \def\cR{{\cal R}}
\def\cS{{\cal S}} \def\cT{{\cal T}} \def\cU{{\cal U}}
\def\cV{{\cal V}} \def\cW{{\cal W}} \def\cX{{\cal X}}
\def\cY{{\cal Y}} \def\cZ{{\cal Z}}

\def\ua{\underline{\alpha}}
\def\ub{\underline{\phantom{\alpha}}\!\!\!\beta}
\def\uc{\underline{\phantom{\alpha}}\!\!\!\gamma}
\def\um{\underline{\mu}}
\def\ud{\underline\delta}
\def\ue{\underline\epsilon}
\def\una{\underline a}\def\unA{\underline A}
\def\unb{\underline b}\def\unB{\underline B}
\def\unc{\underline c}\def\unC{\underline C}
\def\und{\underline d}\def\unD{\underline D}
\def\une{\underline e}\def\unE{\underline E}
\def\unf{\underline{\phantom{e}}\!\!\!\! f}\def\unF{\underline F}
\def\unm{\underline m}\def\unM{\underline M}
\def\unn{\underline n}\def\unN{\underline N}
\def\unp{\underline{\phantom{a}}\!\!\! p}\def\unP{\underline P}
\def\unq{\underline{\phantom{a}}\!\!\! q}
\def\unQ{\underline{\phantom{A}}\!\!\!\! Q}
\def\unH{\underline{H}}

\def\As {{A \hspace{-6.4pt} \slash}\;}
\def\bs {{b \hspace{-6.4pt} \slash}\;}
\def\Ds {{D \hspace{-6.4pt} \slash}\;}
\def\ds {{\del \hspace{-6.4pt} \slash}\;}
\def\ss {{\s \hspace{-6.4pt} \slash}\;}
\def\ks {{ k \hspace{-6.4pt} \slash}\;}
\def\ps {{p \hspace{-6.4pt} \slash}\;}
\def\pas {{{p_1} \hspace{-6.4pt} \slash}\;}
\def\pbs {{{p_2} \hspace{-6.4pt} \slash}\;}
\def\Ps {{P \hspace{-6.4pt} \slash}\;}
\def\Qs {{Q \hspace{-6.4pt} \slash}\;}

\def\Fh{\hat{F}}
\def\Vh{\hat{V}}
\def\Xh{\hat{X}}
\def\ah{\hat{a}}
\def\xh{\hat{x}}
\def\yh{\hat{y}}
\def\ph{\hat{p}}
\def\xih{\hat{\xi}}
\def\psit{\tilde{\psi}}
\def\Psit{\tilde{\Psi}}
\def\tht{\tilde{\th}}
\def\lt{\tilde{\lambda}}
\def\hl{\hat{\lambda}}
\def\hlt{\hat{\tilde{\lambda}}}
\def\llt{\tilde{l}}
\def\At{\tilde{A}}
\def\Qt{\tilde{Q}}
\def\Rt{\tilde{R}}
\def\Nt{\tilde{N}}

\def\at{\tilde{a}}
\def\st{\tilde{s}}
\def\ft{\tilde{f}}
\def\pt{\tilde{p}}
\def\qt{\tilde{q}}
\def\vt{\tilde{v}}
\def\nt{\tilde{n}}

\def\delb{\bar{\partial}}
\def\bz{\bar{z}}
\def\bD{\bar{D}}
\def\bB{\bar{B}}

\def\bk{{\bf k}}
\def\bl{{\bf l}}
\def\bp{{\bf p}}
\def\bq{{\bf q}}
\def\br{{\bf r}}
\def\bx{{\bf x}}
\def\by{{\bf y}}
\def\bR{{\bf R}}
\def\bV{{\bf V}}

\def\d{\delta}\def\D{\Delta}\def\ddt{\dot\delta}
\def\pa{\partial} \def\del{\partial}
\def\xx{\times}
\def\uno{\mbox{1 \kern-.59em {\rm l}}}
\def\trp{^{\top}}
\def\inv{^{-1}}
\def\dag{{^{\dagger}}}
\def\pr{^{\prime}}
\def\lan{\langle}
\def\ran{\rangle}
\def\rar{\rightarrow}
\def\lar{\leftarrow}
\def\lrar{\leftrightarrow}
\newcommand{\0}{\,\!}      
\def\one{1\!\!1\,\,}
\def\im{\imath}
\def\jm{\jmath}
\newcommand{\tr}{\mbox{tr}}
\newcommand{\slsh}[1]{/ \!\!\!\! #1}
\def\vac{|0\rangle}
\def\lvac{\langle 0|}
\def\hlf{\frac{1}{2}}
\def\ove#1{\frac{1}{#1}}
\def\Box{\square}
\def\ZZ{\mathbb{Z}}
\def\CC#1{({\bf #1})}
\def\bcomment#1{}
\def\bfhat#1{{\bf \hat{#1}}}
\def\VEV#1{\left\langle #1\right\rangle}
\newcommand{\ex}[1]{{\rm e}^{#1}} \def\ii{{\rm i}}
\def\rr{{\rm r}} \def\rs{{\rm s}}\def\rv{{\rm v}}
\def\ri{{\rm i}}\def\rj{{\rm j}}
\newcommand{\lrbrk}[1]{\left(#1\right)}
\newcommand{\sfrac}[2]{{\textstyle\frac{#1}{#2}}}

\def\Li{{\rm Li}_2}

\font\mybb=msbm10 at 12pt
\def\bb#1{\hbox{\mybb#1}}

\font\myBB=msbm10 at 18pt
\def\BB#1{\hbox{\myBB#1}}

\begin{document}

\begin{flushright}
QMUL-PH-12-01 \\
WIS/03/12-JAN-DPPA
\end{flushright}

\vspace{8pt}

\begin{center}

{\Large \bf Analytic two-loop form factors in $\mathcal{N}=4$  SYM  }

%
\vspace{16pt}

{\mbox {\bf  Andreas Brandhuber$^{a, b}$,  Gabriele Travaglini$^{a}$ and   Gang Yang$^{c}$}}%
\footnote{
{\tt  \{ \tt \!\!\!a.brandhuber, g.travaglini\}@qmul.ac.uk, gang.yang@desy.de}
}


\begin{quote}
{\small \em
\begin{itemize}
\item[\ \ \ \ \ \ $^a$]
\begin{flushleft}
Centre for Research in String Theory\\
School of Physics and Astronomy\\
Queen Mary University of London\\
Mile End Road, London E1 4NS, United Kingdom
\end{flushleft}
\item[\ \ \ \ \ \ $^b$]
Department of Particle Physics and Astrophysics\\
Weizmann Institute of Science,
Rehovot 76100, Israel
\item[\ \ \ \ \ \ $^c$]
II. Institut f\"ur Theoretische Physik, Universit\"at
Hamburg\\ Luruper Chaussee 149, D-22761 Hamburg, Germany

\end{itemize}
}
\end{quote}


\vspace{60pt} {\bf Abstract}
\end{center}

\noindent
We derive a compact expression for the  three-point MHV form factors of half-BPS operators in $\mathcal{N}=4$ super Yang-Mills at two loops. The main tools of our calculation are generalised unitarity applied at the form factor level, and the compact expressions for
supersymmetric tree-level form factors and amplitudes entering the cuts.
We confirm that  infrared divergences exponentiate as expected, and that collinear factorisation is entirely captured by an ABDK/BDS ansatz. Next, we construct  the two-loop remainder function obtained by subtracting this ansatz from the full two-loop form factor and compute it numerically. Using symbology, combined with various physical constraints and symmetries, we find a unique solution for its symbol. With this input we construct a remarkably compact  analytic expression for the remainder function, which contains only classical polylogarithms, and compare it to our numerical results. Furthermore, we make the surprising observation that our remainder is equal to the maximally transcendental piece of a closely related two-loop amplitude in QCD.

\setcounter{page}{0}
\thispagestyle{empty}
\newpage


\setcounter{tocdepth}{4}
\hrule height 0.75pt
\tableofcontents
\vspace{0.8cm}
\hrule height 0.75pt
\vspace{1cm}

\setcounter{tocdepth}{2}


\setcounter{footnote}{0}

 \section{Introduction}

The last years have witnessed
dramatic progress in our understanding of seemingly unrelated physical quantities in $\cN=4$ supersymmetric Yang-Mills theory (SYM), in particular scattering amplitudes,   Wilson loops and correlations functions.  One important lesson is that these objects are actually related in unexpected ways, which has led to major conceptual insights, vastly improved techniques to calculate them and the discovery of new structures and symmetries.

In this context it is  natural to consider simple, interesting  generalisations of the  quantities mentioned above, such as form factors.
In a gauge theory, one considers the overlap of a state created by a gauge-invariant operator,
$\cO(x) | 0 \ran$, with a multiparticle state $\lan 1 \cdots n |$. The form factor is then defined as
\beq
\int\!\!d^4x \, e^{-iqx} \, \lan 1 \cdots n | \cO (x) |0\ran \ =   \ \delta^{(4)} (q - \sum_{i=1}^n p_i ) \, F(1, \ldots , n )\ ,
\eeq
where
$  F(1, \ldots , n ) := \lan 1 \cdots n | \cO (0) |0\ran$.

For $n=2$, the latter  is the celebrated Sudakov form factor,  which is a famous example of exponentiation -- a phenomenon discovered
more recently in scattering amplitudes of $\cN=4$ SYM \cite{abdk, bds}, and is intimately linked to scattering amplitudes,  whose universal infrared divergences are captured by  Sudakov form factors. An $n$-point form factor can also be viewed as describing the decay of a state created by the gauge-invariant operator $\cO$ into an $n$-particle  state.  For $n=2$ the kinematic dependence is trivial,  and the form factor can be expressed in terms of the cusp and collinear anomalous dimensions. In planar $\cN=4$ SYM,  the former is known to all orders in the gauge coupling \cite{BES}.

In this paper we will  mainly be concerned with form factors with $n>2$,  which have highly non-trivial kinematic dependence. While  form factors share several properties  with  scattering amplitudes,
such as soft and collinear factorisation,  they also display interesting and important differences. This is due to the insertion of a gauge-invariant operator, which  carries momentum that  can be injected at any location in Feynman diagrams, and is thus not restricted by the colour ordering of the external, on-shell particles.
In practice this leads to a more complicated colour management and to the appearance of non-planar loop integral topologies,  even if we restrict ourselves to the planar limit, as we do in this paper.
Furthermore, this also implies that the dual conformal (super)symmetry of planar amplitudes in $\cN=4$ SYM is not a property of form factors.

Form factors at strong coupling in $\cN=4$ SYM were  recently studied in  \cite{Alday:2007he,mz}.  The particular choice of the operator considered in those papers is immaterial, as long as its conformal dimension is small compared to $\sqrt{\lambda}$.
Specifically, in \cite{mz}  form factors were analysed in great detail and  the
explicit expression of the four-point  form factor  was determined in the case of $(1+1)$-dimensional kinematics.  This  is the first example of a nontrivial form factor remainder  at strong coupling.
In \cite{bsty, harmony} we have begun a systematic investigation of form factors of bilinear  half-BPS operators at weak coupling, following a pioneering paper of Van Neerven \cite{VN1}.%
\footnote{A parallel investigation was carried out in   \cite{Bork:2010wf, Bork:2011cj}.}
In particular, in \cite{bsty} we determined the explicit one-loop expressions for these  form factors for an arbitrary  number of external particles with an MHV helicity configuration.

In this paper we will focus on the two-loop calculation of the three-point form factor. In QCD, similar quantities
have been calculated  at one  \cite{Schmidt:1997wr} and, more recently,  two loops \cite{Nigel, kkt} using Feynman diagrams. These form factors are phenomenologically important since they are related to the scattering of $e^+ e^- \to 3 \, \mathrm{jets}$ and $H\to 3 \, \mathrm{jets}$.
Here we will use instead (generalised) unitarity \cite{bddk,fusing,bdkgen, bcfgen} applied directly at the level of the form factor \cite{bsty} in order to determine the two-loop, three-point  form factor from its cuts.

In Section 2 we will set the stage by presenting the details of a simpler calculation, namely that of the Sudakov form factor at two loops. Section 3 describes the calculation of our three-point form factor at two loops from generalised cuts. We obtain a compact result, presented in \eqref{result}, expressed in terms of planar and non-planar two-loop integral functions. Most (but not all) of the analytic expressions for these functions have been derived in \cite{gr99, grp, grnp}. For this reason, we have resorted to numerical methods in order to collect data about this form factor which can then be compared to analytic expressions. The main tool here is the Mellin-Barnes representation of higher-loop integrals (see for example \cite{smirnov}), and their evaluation using the powerful numerical algorithm of \cite{czakon}.

In order to present the two-loop form factor in an efficient way, we construct in Section 4 a remainder function very much in the way as for the case of scattering amplitudes. At two loops, this remainder is
\beq
\cR_n^{(2)} \ := \    \cG^{(2)}_n(\e )\, - \, {1\over 2} \big( \cG^{(1)}_n (\e) \big)^2 -  f^{(2)} (\e) \cG^{(1)}_n  ( 2 \e ) \, - \,  C^{(2)}
\, + \cO (\e ) \
\ ,
\eeq
where $\cG^{(L)}_n$ is the colour-stripped $L$-loop form factor divided by its tree-level expression, $C^{(2)}$ is an $n$-independent constant and $ f^{(2)}(\eps)$ contains the cusp and collinear anomalous dimensions.
One important reason to introduce this remainder is that, as we show explicitly in Section 4, it has the correct collinear limits. More precisely, if $C^{(2)}$ is chosen appropriately then
$\cR_2^{(2)}=0$ and in a simple collinear limit
$\cR_{n}^{(2)} \to \cR_{n-1}^{(2)}$.
In other words, the ABDK/BDK ansatz for the form factor captures correctly the collinear behaviour of the two-loop form factor. The remaining task is then to determine this remainder function which is a function of scale invariant ratios of Mandelstam variables only.

Our main weapon to attack this remainder analytically is the use of symbols \cite{symbol}. This is a very powerful concept that allows to re-express complicated identities between polylogarithmic functions in terms of simple algebraic identities. It was used in \cite{symbol} to rewrite  the Wilson loop six-point remainder function found in \cite{vittorioetal}
in a very compact form. Because of the amplitude/Wilson loop duality \cite{am,dks,bht}, this remainder is also equal to the six-point MHV amplitude remainder at two loops, as was confirmed numerically in \cite{seven,dhkshex}. Recent interesting applications of the symbol to the construction of various remainder functions in $\cN=4$ SYM were presented in \cite{pulling, CaronHuot:2011ky, Dixon:2011pw, Heslop:2011hv, Dixon:2011nj, Prygarin:2011gd}. 

As we will discuss in Section 4, we can impose several physical constraints, such as collinear limits and various requirements on the first, second and last entry of the symbol, directly at the level of the symbol. In the present instance, we find that the various physical properties are so stringent that the symbol of the two-loop remainder function at three points is determined uniquely; we present its explicit expression  in \eqref{nice}. An important point we would like to anticipate is that the symbol of the three-point form factor at two loops depends on a very restricted set of variables, as we discuss in detail in Section 4. Furthermore, it obeys a particular symmetry condition  \cite{symbol} which guarantees that the corresponding function can be written in terms of classical polylogarithms up to degree four only.
Using this information, we will construct a surprisingly simple  analytic expression for the complete remainder function, shown in \eqref{beauty}. We note that beyond-the-symbol ambiguities can easily be fixed by imposing correct behaviour in collinear limits.
Within errors,  our result is in good agreement with our numerical results obtained for various kinematic points.

We also observe an interesting relation between  the symbol of the three-point form factor remainder  and a piece  of the six-point MHV amplitude remainder of \cite{symbol} upon appropriate identification of the kinematic variables. This is reminiscent of a result found in \cite{mz}, where the four-point form factor in $(1+1)$-dimensional kinematics was evaluated at strong coupling and expressed in terms of the eight-point MHV amplitude remainder  in $(1+1)$-dimensional kinematics.

We will conclude by establishing an intriguing relation between our remainder  and  the finite remainder function (defined with a prescription introduced by Catani  \cite{magician})  of the two-loop amplitude $H \to ggg$ in QCD, a quantity that was recently presented in \cite{Nigel, kkt} in the large top mass limit. Interestingly, if we translate the latter
into a  ABDK/BDS remainder function and extract the terms of highest transcendentality, which is
four at two loops, we find a function that has the same symbol as the remainder of the
three-point form factor calculated in this paper. In fact, this observation provides us with an alternative way to find the analytic $\cN\!=\!4$ SYM remainder, although our expression is much more compact -- specifically, the  analytic formula obtained pursuing this alternative strategy  contains also two-dimensional harmonic polylogarithms (2dHPL's) \cite{grp,grnp} which, as we mentioned, are  not present in  our expression \eqref{beauty}. Remaining beyond-the-symbol ambiguities can again be fixed using collinear limits and symmetry.
We have checked by explicit comparison that these two representations of the remainder  are in complete agreement.

To our knowledge, this is the first occurrence of the principle of maximal transcendentality  \cite{maxtrans}, which relates the maximally transcendental part of a quantity in QCD to the same quantity in planar $\cN=4$ SYM, in an observable  with non-trivial kinematic dependence. It would be  interesting to find more examples of this, and understand  whether this observation can be used to simplify the calculation and expressions of other scattering amplitudes in QCD.


\section{Warm up: the Sudakov form factor at two loops}

In this section we present a very simple, unitarity-based derivation
of the two-loop Sudakov form factor $F(q^2) :=  \lan \phi_{12} (p_1)
\phi_{12} (p_2) | \Tr \big(\phi_{12}\phi_{12}\big) (0) | 0 \ran$,
where $q:= p_1 + p_2$. This quantity was  calculated at one and two
loops in $\cN=4$ SYM  in \cite {VN1} using unitarity applied to
Feynman diagrams \cite{VN2}, see also  \cite{bsty, moch, camille},
and recently at three loops in \cite{Gehrmann:2011xn}. Here we will
use an approach based on unitarity directly applied at the  level of
the form factor. This approach was  introduced in \cite{bsty} and is
a straightforward  generalisation of the unitarity-based method of
\cite{bddk, fusing}.

There is only one kinematic channel, the $q^2$ channel, hence  it is
sufficient to consider two-particle cuts, and lift the cut integral
to a full $D$-dimensional integral, as in \cite{collproof}.  There
are two distinct configurations contributing to such a  cut:  in the
first one,  a tree-level, two-point form factor and a one-loop
amplitude enter the cut, whereas in the second one  we have a
one-loop two-point form factor, and a tree-level four-point
amplitude. Let us focus on  the first possibility. Furthermore, for
the sake of illustration,  we will perform the calculation in two
different ways: firstly, using the conventional representation of
the one-loop amplitude entering the cut  in terms of fundamental
colour generators, and then using its representation in terms of
adjoint generators  \cite{dddm}.

%
%
\begin{center}
\begin{figure}[t]
\centerline{\includegraphics[height=3cm]{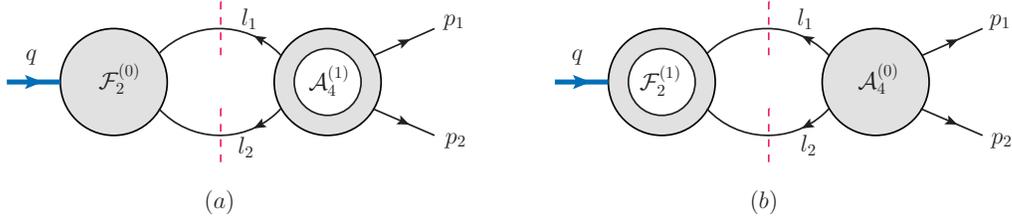} } \caption{\it
The two two-particle cuts contributing to the Sudakov form factor.}
 \label{2loopcut-2p}
\end{figure}
\end{center}
%
%

\subsection{Calculation with fundamental generators}
The first quantity entering the cut is the four-point one-loop amplitude, whose expression in terms of fundamental colour generators is given in \eqref{cd} with $n=4$. This expression has to be convoluted with a tree-level two-point form factor, simply given by $\delta^{a_{l_1} a_{l_2}}$, see Figure~\ref{2loopcut-2p}{\it a}. We focus first on the contribution from the planar amplitude, i.e.~the first line in \eqref{cd}. There are six possible permutations to consider, which give rise to the single-trace structures%
\footnote{To keep the notation simple we define $\Tr(1,2,l_1, \cdots) := \Tr (T^{a_1} T^{a_2} T^{a_{l_1}}\cdots)$ etc.}
\beqa
\label{trstr}
&&\Tr(1,2,l_1, l_2), \quad \Tr(1,2,l_2, l_1), \quad \Tr(1,l_1, l_2, 2), \quad  \Tr(1,l_2, l_1, 2), \nonumber \\ &&\Tr(1,l_1, 2, l_2), \quad \Tr(1,l_2, 2, l_1).
\eeqa
When contracting with the tree form factor $\lan \phi_{12}^{a_{l_1}} (l_1) \phi_{12}^{a_{l_2}} (p_2) | \Tr \big(\phi_{12}\phi_{12}\big) (0) | 0 \ran = \delta^{a_{l_1} a_{l_2}} $, and using $\delta^{ab} T^{a}_{ij} T^b_{lm} =  \delta_{im} \delta_{jl}$ and $\delta^{ab} T^{a}_{ij} T^b_{jm} = N \delta_{im}$
we see that only the first line of \eqref{trstr} is leading in colour. The four traces give an identical result, namely $N \Tr(a_1 a_2) = N^2 \delta^{a_1 a_2}$.
The contribution to the cut from the planar one-loop amplitude is then
\beq
N^2 \delta^{a_1 a_2} \Big[ A^{(1)}_{4; 1} (1, 2, l_1, l_2) \, + \,  A^{(1)}_{4; 1} (1, 2, l_2, l_1) \, + \, A^{(1)}_{4; 1} (2,1, l_1, l_2)\, + \, A^{(1)}_{4; 1} (2,1,  l_2, l_1)\Big]\, .
\eeq
We now consider the contribution from the non-planar part of the one-loop amplitude (see section A for details). The COP for $\{\alpha\} = \{ 2,1\}$ and $\{\beta\} = \{ 3,4 \}$ (corresponding to $c=3$ in \eqref{ccc}) are
 \beq
 (2,1,3,4), \quad ( 2,3,1,4), \quad (3,2,1,4), \quad (1,2,3,4), \quad (1,3,2,4), \quad (3, 1, 2, 4),
 \eeq
and the non-planar one-loop piece (second line of \eqref{cd}) is
 \beqa
 \cA_{\rm NP}^{(1)}(1, \ldots 4) & = &
 \Tr(1,2) \Tr(l_1, l_2) A_{4; 3}^{(1)}(1,2,l_1, l_2) + \Tr(1,l_1) \Tr(2, l_2) A_{4; 3}^{(1)}(1,l_1, 2,l_2) \nonumber \\
 &+&
  \Tr(1, l_2) \Tr(2, l_1) A_{4; 3}^{(1)}(1,l_2, 2,l_1)
  \ ,
  \eeqa
  with
\beqa
A_{4; 3}^{(1)} (1,2,l_1, l_2) &=& A_{4; 1}^{(1)}( 2,1,l_1, l_2) + A_{4; 1}^{(1)} ( 2, l_1, 1, l_2) + A_{4; 1 }^{(1)} (l_1, 2, 1, l_2) \nonumber \\
&+& A_{4; 1}^{(1)}(1, 2, l_1, l_2) + A_{4; 1}^{(1)}(1, l_1, 2, l_2) + A_{4; 1}^{(1)} (l_1, 1, 2, l_2) \, .
 \eeqa
 Contracting with the tree-level form factor, we see that the leading structures in colour are of the form
 $\delta^{a_{l_1} a_{l_2}} \Tr(l_1 l_2) \Tr(1 2) = N^2 \delta^{a_1 a_2} $.
 Collecting terms, we obtain a two-loop cut integrand
 \beq
 \label{ffc1}
\left.
 F^{(2) a_1 a_2}(q^2)
\right|_{q^2-\mathrm{cut}} \ = \ 2\, N^2 \delta^{a_1 a_2}\
 \int\!\!d{\rm LIPS} (l_1, l_2; q)\
 \Big[
 4\,  A_{4; 1}^{(1)} (1,2,l_1, l_2) \, + \,  A_{4; 1}^{(1)}(1,l_1, 2, l_2)
 \Big]
 \ ,
 \eeq
where the Lorentz invariant phase space measure in dimensional regularisation is
\beq
d{\rm LIPS}(l_1, l_2; q) := d^D l_1\, d^D l_2 \, \delta^+ (l_1^2)  \delta^+ (l_2^2) \delta^D (l_1+ l_2  + q)\ ,
\eeq
with $D=4-2\eps$.
The one-loop  component amplitudes appearing in \eqref{ffc1}
are given, to all orders in the dimensional regularisation parameter $\eps$ by
\beqa
 A_{4; 1}^{(1)} (1,2,l_1, l_2)&=& A^{(0)}\big(  \phi_{12} (p_1),  \phi_{12} (p_2),  \phi_{34} (l_1),  \phi_{34} (l_2) \big) \, F^{\rm 0m}(1,2,l_1, l_2)
 \nonumber \\
 A_{4; 1}^{(1)} (1,l_1, 2, l_2)&=& A^{(0)}\big(  \phi_{12} (p_1),  \phi_{34} (l_1), \phi_{12} (p_2),    \phi_{34} (l_2) \big) \, F^{\rm 0m}(1,l_1, 2, l_2)
\eeqa
where $F^{\rm 0m}(1, \ldots, 4) = s_{12} s_{23} I^{\rm 0m}(1, \ldots, 4)$ is a zero-mass box function. The tree amplitudes $A^{(0)}\big(  \phi_{12} (p_1),  \phi_{12} (p_2),  \phi_{34} (l_1),  \phi_{34} (l_2) \big)$ and $A^{(0)}\big(  \phi_{12} (p_1),  \phi_{34} (l_1), \phi_{12} (p_2),    \phi_{34} (l_2) \big)$
can be extracted from Nair's superamplitude \cite{Nair}
\beq
\label{superMHV}
\cA_{\rm MHV} :=  g^{n-2} \,  (2 \pi)^4 \delta^{(4)}  \big(\sum_{i=1}^n \l_i \lt_i \big) \, \delta^{(8)} \big(\sum_{i=1}^n \l_i \eta_i \big) \, \prod_{i=1}^n {1\over \lan i i+1 \ran}
\ ,
\eeq
where
$\l_{n+1} \equiv \l_1$, giving%
\footnote{Note that $s_{12} s_{2l_1} A^{(0)}\big(  \phi_{12} (p_1),  \phi_{12} (p_2),  \phi_{34} (l_1),  \phi_{34} (l_2) \big) =
s_{1l_1} s_{l_12}A^{(0)}\big(  \phi_{12} (p_1),   \phi_{34} (l_1),  \phi_{12} (p_2), \phi_{34} (l_2) \big)$, which simplifies the cut algebra.}
\beqa
\label{ffc2}
A^{(0)}\big(  \phi_{12} (p_1),  \phi_{12} (p_2),  \phi_{34} (l_1),  \phi_{34} (l_2) \big) &=& { \lan l_1 l_2 \ran  \lan12\ran\over
\lan l_2 1 \ran  \lan 2 l_1 \ran } \ ,
\\
A^{(0)}\big(  \phi_{12} (p_1),   \phi_{34} (l_1),  \phi_{12} (p_2), \phi_{34} (l_2) \big)&=& { \lan 1 2 \ran^2  \lan l_1 l_2\ran^2\over
\lan 1 l_1 \ran  \lan l_1  2 \ran  \lan 2 l_2 \ran \lan l_2 1 \ran } \ .
\eeqa
On the cut one easily finds
\beq
{\lan 12\ran \lan l_1 l_2 \ran \over \lan 2 l_1\ran  \lan l_2 1\ran} = - {q^2 \over (l_2 + p_1)^2}
\ .
\eeq
Using this relation we easily arrive at
\beq
F^{(2) a b }(q^2) \ = \  N^2 \, \delta^{ab} \, F^{(2)}(q^2)
\ ,
\eeq
with
\beq
 \label{ffsuda}
 F^{(2)}(q^2)  \ = \  2 \Big[ 4\, {\rm LT} (q^2, \eps)  \, + \,  {\rm CT} (q^2, \eps)\Big]
 \ ,
 \eeq
where  the two-loop ladder and crossed triangle, $ {\rm LT} (q^2, \eps)$,  ${\rm CT }(q^2, \eps) $ respectively,
are given by \cite{Gonsalves:1983nq, VN2, Kramer:1986sr, Gehrmann:2005pd, smirnov}%
\footnote{In the following formulae we actually divide  these functions  by a power of $q^2$ per loop. }
 \beqa
{\rm LT} (q^2, \eps  )  &=&  (-q^2)^{-2\eps}
 e^{ 2  \gamma \eps }
 \bigg\{
   {1\over \eps}  \bigg[  {1\over 2 \eps } G (2, 2 ) G_3 (2 + \eps, 1, 1)\\
  &&  -
      G(2, 1)  \Big[ {1\over \eps}  G_3 ( 2, 1, 1 + \eps)  + G_3 (1, 1, 1) \Big] \bigg]  \bigg\}
      \nonumber \\
      &=&
     (-q^2)^{-2\eps}  \left[ {1\over 4 \eps^4} + { 5 \pi^2\over 24 \eps^2}  + {29\over 6 \eps}  \zeta_3  + {3\over 32} \pi^4 + \cO (\eps) \right]
     \ ,
   \nonumber    \\
{\rm CT}(q^2, \eps) &= &  (-q^2)^{-2\eps} \left[{1\over \eps^4} - {\pi^2\over \eps^2}   - {83\over 3 \eps}  \zeta_3  - {59\over 120}  \pi^4 \, + \, \cO (\eps) \right]
\ ,
\eeqa
where
\beqa
G(x,y) & =  &
{ \Gamma ( x + y +  \eps - 2)  \Gamma (
   2 - \eps - x )  \Gamma (  2 - \eps - y )  \over
     \Gamma (x)  \Gamma (y)  \Gamma ( 4 - x - y - 2 \eps)  }
\ ,
\\
G_3 ( x, y, z)  &=&
{ \Gamma ( 2- x - z - \eps )  \Gamma ( 2- y - z - \eps  ) \Gamma (
   -2+x + y + z + \eps  )
 \over
      \Gamma ( x ) \Gamma (y)  \Gamma (4 -x - y - z - 2 \eps  )   }
\ .
\eeqa
Going through the same step for the cut in Figure~\ref{2loopcut-2p}{\it b} easily confirm the result we have derived here.


\subsection{Calculation with adjoint generators}
The same calculation outlined in the previous section can be described in a very efficient way using the representation of one-loop amplitudes in terms of adjoint generators  \cite{dddm}  described in \eqref{vitto}. We consider again a two-particle cut with a tree-level form factor and a one-loop four-point amplitude
\beq
\mathcal{A}^{(1)}(1, 2, l_1, l_2)\ = \  \sum_{\sigma\in S_4 / (\mathbb{Z}_4 \times \mathcal{R})} {\rm Tr} (F^{a_{\sigma_1}}\cdots F^{a_{\sigma_4}})\, A_{4; 1}^{(1)}  (\sigma_1, \ldots , \sigma_4) \, ,
\eeq
 where $(\sigma_1, \ldots , \sigma_4)$ are permutations of $(1, 2, l_1, l_2)$. The sum contains three terms, associated with the orderings
 $(1, 2, l_1, l_2)$,  $(1, 2, l_2, l_1)$
 $(1, l_1, 2, l_2)$. On the cut, we contract the colour labels $a_{l_1}$, $a_{l_2}$ of the loop legs with a $ \delta^{a_{l_1}a_{l_2}} $ from the two-point tree-level  form factor, as already described in the previous section. Doing this,
 the first two, and the last colour orderings will give rise to  the following two colour traces
 \beq
 \label{44}
 \delta^{a_{l_1}a_{l_2}} \Tr (F^{a_1} F^{a_2} F^{a_{l_1}} F^{a_{l_2}}) \, , \qquad   \delta^{a_{l_1}a_{l_2}}  \Tr (F^{a_1} F^{a_{l_1}} F^{a_2} F^{a_{l_2}})
 \ ,
\eeq
respectively, where%
\footnote{Because of the slightly unconventional normalisation of the adjoint generators, we have
 $\Tr (F^a F^b) = 2 N \delta^{ab}$, and  $(F^a F^a)_{bc} = 2N \delta_{bc}$.
 This different normalisation for the adjoint generators is responsible for the appearance of factors of 4 in \eqref{44}. }
\beq
\label{1/2}
\Tr (F^{a_1} F^{a_2} F^a F^a) \, = \,  4\, N^2 \delta^{a_1 a_2} \, , \qquad
\delta^{ab} \Tr (F^{a_1} F^{a} F^{a_2} F^a)\, = \,4\,  {N^2\over 2} \delta^{a_1 a_2}
\ .
\eeq
Taking into account these relations, one instantly arrives at \eqref{ffc1}.

Before closing this section, we also quote two useful formulae when dealing with adjoint traces, namely
\beq
\Tr (F^a A F^a B) \ = \  \Tr (F^a A^t F^a B) + \Tr (F^a A) \Tr (F^a  B)
\ ,
\eeq
which, in the case where $A^t = - A$, simplifies to
\beq
\Tr (F^a A F^a B) \ = \ {1\over 2} \Tr (F^a A) \Tr (F^a  B)   \qquad {\rm if } \quad A^t = - A
\, .
\eeq
This explains the factor of $1/2$ in the second relation in \eqref{1/2}.


\section{Generalised unitarity for planar form factors}

In this section we compute higher-point form factors using  generalised
unitarity \cite{bdkgen, bcfgen}. The explicit calculation
will focus on the MHV three-point two-loop form factor,
\be
\label{defineff3}
F_3^{\mathrm{MHV},(2)}(1,2,3) :=  \lan \phi_{12} (p_1) \phi_{12} (p_2) g^+(p_3) | \Tr \big(\phi_{12}\phi_{12}\big) (0) | 0 \ran\ ,
\ee
for which we will now derive the complete result. Recall that the operator $\Tr \big(\phi_{12}\phi_{12}\big)$ is a half-BPS operator and is the lowest component of the stress-tensor multiplet. Hence, we expect the result to be the same if this operator is replaced by any other operator in the stress-tensor multiplet \cite{harmony}. A different choice of operator usually requires a different choice of external states as well.
We also notice that in $\cN=4$ SYM there is nothing special about the particular choice of helicities of the external states in \eqref{defineff3} -- any other choice with the same total helicity and $R$-charge are possible.
In fact, we could have worked in a more invariant way by using  supersymmetric form factors as introduced in \cite{harmony},  and in the case of MHV form factors all helicity dependence factors out into the MHV tree-level (super) form factor,
\be
F_3^{\mathrm{MHV},(2)}(1,2,3) =
F_3^{\mathrm{MHV},(0)}(1,2,3) \, \mathcal{G}^{(2)}_3(1,2,3) \ ,
\ee
where the helicity-blind function $\mathcal{G}^{(2)}_3(1,2,3)$
depends only on the momenta through $s_{12}:= 2 p_1 \cdot p_2$,
$s_{23}:= 2 p_2 \cdot p_3$, $s_{31}:= 2 p_3 \cdot p_1$,  and $q^2 :=
s_{12}+s_{23}+s_{31}$; finally, $F_3^{\mathrm{MHV},(0)}(1,2,3)$ is
the tree-level (super) form factor. For the particular helicity configuration in \eqref{defineff3} this tree-level form factor
is
$\lan 12 \ran/ (\lan 2 3 \ran \lan 3 1 \ran)$.
It is easy to see that in all
calculations this tree-level piece factors out naturally,  and
therefore the goal is to determine
$\mathcal{G}^{(2)}_3(1,2,3)$. We mention that this
ratio function is also shared by the NMHV three-point form factor
defined by factoring out the NMHV tree-level form factor, due to the
chiral symmetry of three-point case \cite{harmony}.
We also note  in passing that three-point form factors with adjoint
fields always have a colour factor proportional to the structure
constants $f^{a_1 a_2 a_3}$.

We begin by reconsidering the cut we are familiar with from the two-particle two-loop
form factor, namely the two-particle cut in the
$q^2$ channel, where the cut integrand is given as
the product of a tree-level two-point form factor and a one-loop
amplitude. As we have seen in the two-point case, the double-trace
non-planar amplitudes also contribute to planar form factors because
of the colour index contraction
\be  {\rm Tr}(T^{a_{l_1}} T^{a_{l_2}})\, [{\rm Tr}(T^{a_{l_1}}
T^{a_{l_2}})\, {\rm Tr}(T^{a_1} \cdots T^{a_n})]  = N^2 \, {\rm
Tr}(T^{a_1} \cdots T^{a_n}) \,. \ee
Using the general formulae for the one-loop amplitudes
\eqref{cd}-\eqref{ccc}, we immediately obtain the corresponding cut integrand of the planar
$n$-point form factor,
\bea \label{ffcnp} F_n^{(2)}(q^2) \Big|_{q^2-\mathrm{cut}} & = &
\int\!\!d{\rm LIPS} (l_1, l_2; q) \, \sum_{\sigma\in S_n}\,
\Big[4 \, A_{n;1}(l_1, l_2, \sigma(1), \ldots \sigma(n)) \\
&& \qquad~ + \, 2 \sum_{i=1}^{\lfloor n-1 \rfloor/2} A_{n;1}(l_1,
\sigma(1),\ldots, \sigma(i), l_2, \sigma(i+1), \ldots, \sigma(n))
\nonumber\\ && \qquad~ + \, A_{n;1}(l_1, \sigma(1),\ldots,
\sigma(n/2), l_2, \sigma(n/2+1), \ldots, \sigma(n)) \Big]
\, ,  \nonumber \eea
where the last line contributes only when $n$ is even. In particular,
for three-point form factors we have
\bea \label{ffc3p} F_3^{(2)}(q^2) \Big|_{q^2-\mathrm{cut}} & = &
\int\!\!d{\rm LIPS} (l_1, l_2; q) \\
&& \sum_{\sigma\in S_3}\, \Big[4 A_{5;1}(l_1, l_2, \sigma(1),
\sigma(2), \sigma(3)) + 2 A_{5;1}(l_1, \sigma(1), l_2, \sigma(2),
\sigma(3)) \Big] ~. \nonumber \eea
Note that in the above presentation we have fixed the order of
the external on-shell states,  and stripped off the colour factor
$ N^2 {\rm
Tr}(T^{a_1} T^{a_2} \ldots T^{a_n})$.  We call this quantity a
colour-stripped form factor, or colour-ordered form
factor. The complete planar form factor can be expressed in terms
of these colour-stripped form factor times the appropriate colour factor,
and summing
over non-cyclic permutation of external on-shell states.
It is this
colour-ordered form factor that we will focus on in the later
calculation.

We would also like to point out  that colour-ordered form factors  can also be
computed directly using colour-stripped Feynman rules, which are simple
generalisation of the well-known colour-stripped Feynman rules for
amplitudes.  The relevant diagrams are planar diagrams  -- in the sense
that  we fix the ordering of the external on-shell particles, while
the off-shell momentum $q$ of the gauge-invariant operator can
appear in any possible position. Hence, in momentum space,  diagrams
can look non-planar, while in fact in colour-space they are planar.

We will now focus on the computation of the three-point form factor. Unlike the
two-point case, it is technically more difficult to lift the two-particle $q^2$ cut directly to the two-loop
form factor, because of the presence of the one-loop five-point
amplitude \eqref{ffc3p}. However, we have used these cuts and two-particle cuts in other channels as important cross-checks of our final result.

For $n$-point form factors,  not much is known about the set of two-loop integral functions
appearing (except for $n=3$ \cite{grp,grnp}),  and since dual conformal symmetry is not a feature of form factors it cannot be used to reduce the set of allowed integrals. Furthermore,
planar form factors, as already seen in the case of the Sudakov form factor, contain non-planar integral topologies.
Therefore, we follow a two-step procedure that does not rely on the knowledge of a particular basis of integral functions: in the first step we use double two-particle
cuts to identify a basis. This set of cuts leads to sufficiently simple integrands so that their tensor reductions can be performed with relative ease and a simple set of integral functions appears.
In this step we determine all possible integral topologies together with their coefficients, but are left with ambiguities in the numerators of the form $\sim l_i^2$, where $l_i$ is one of the cut momenta. In the second step we perform   more intricate and  stringent triple cuts to fix any remaining ambiguities. One additional complication here is that sums over the helicities of the internal particles have to be performed. However, this can be dealt with elegantly using superamplitudes and super form factors \cite{harmony}.
This procedure leads to a unique answer that passes all consistency checks.

\begin{figure}[t]
\centerline{\includegraphics[height=6cm]{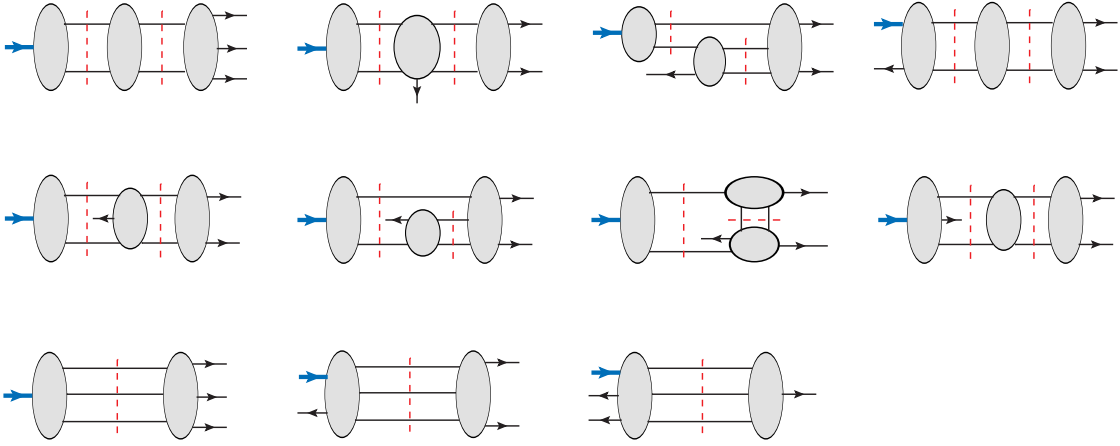} } \caption{\it Double
two-particle cuts and three-particle cuts of three-point form
factor.} \label{cuts}
\end{figure}

More concretely, following the strategy outlined above
we determined the integrals by considering the set of double two-particle
cuts and three-particle cuts shown in  Figure \ref{cuts}. These
cuts are sufficient to determine the three-point form factor
completely. We have also considered various two-particle cuts with
one-loop form factor times tree-level amplitudes, and tree-level
form factor times one-loop amplitudes which provided important additional cross checks,
but did not lead to new integral functions.

%
%
%
%

We now  provide some more details on the procedure we followed.
Starting from the two-particle cut expression \eqref{ffc3p}, one can
apply a further two-particle cut to the one-loop five-point
amplitudes. The cut integrand is then given by the product of
a two-point tree-level  form factor and two tree amplitudes. We also
consider the cuts which are given by a three-point tree form factor
and two tree amplitudes. These types of cuts are depicted in the
first two lines of Figure \ref{cuts}. The cut integrands are simple
enough to perform the necessary tensor reduction directly. In this
way,  we find a set of simple integral functions with simple
coefficients containing all integrals that appear in the final
answer,  given in Figure \ref{fig1} %
\footnote{There are also additional non-planar
integrals arising from the reduction of the double trace term in (\ref{ffc3p}),
some of which are not allowed by the colour structure. However, these are
all canceled after performing  cyclic summation, and do not appear in the final
result. }.

However, we are left with certain ambiguities due to $l_{1,2}^2$ terms
in the numerator of integrals such as $DBox$ and $NBox$ in
Figure \ref{fig1}, where $l_{1,2}$ correspond to cut propagators. Such terms
are not detected in the double two-particle cuts considered. Besides, there are
also integrals which are not detected by these double two-particle
cuts. Both problems can be fixed by considering three-particle
cuts.

The three-particle cuts on their own involve several  integral topologies.%
\footnote{Integrals which are simple products of one-loop integrals
are not detected by three-particle cut, but they are ruled out by
the double two-particle cuts.}
The cuts we have considered are
shown in the third line of Figure \ref{cuts}, and  involve up to
six-point NMHV amplitudes and five-point NMHV form factors. The cut
integrands are therefore much more complicated compared to double
two-particle cuts, which makes it much harder to perform the tensor
reduction directly in order to obtain a set of simple integral functions.
That is why we chose to use first double two-particle cuts to write down
an ansatz,  which we then verify and refine using three-particle cuts.

For these checks we do not need the reduction of the triple-cut
expressions  as we can make analytic comparisons of the {\it
integrands} arising  from the triple cut and those coming from the
ansatz directly, by choosing a basis for spinors and expressing both
integrands in this basis.
%
%
\begin{figure}[t]
\centerline{\includegraphics[height=2.5cm]{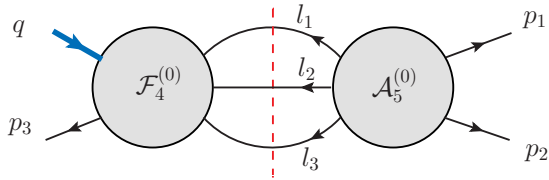} }
\caption{\it A particular three-particle cut of the three-point form factor.}
\label{triple-cut}
\end{figure}
%
%
%

To be more explicit, let us consider the particular three-particle cut shown in
Figure \ref{triple-cut}.  The cut integrand is given as a product of a
four-point tree form factor,  and a five-point tree amplitude. There
are two kinds of contributions to the cut integrand, depending on which one is
next-to-MHV:
\bea && \int d^4 \eta_{l_1} d^4 \eta_{l_2} d^4 \eta_{l_3} \Big[
{\cal F}_4^{\rm MHV,(0)}(-l_1, -l_2, -l_3, 3) \, {\cal A}_5^{\rm
NMHV,(0)}(1, 2, l_3, l_2, l_1) \nonumber\\ && \qquad \qquad \qquad
\quad + {\cal F}_4^{\rm NMHV,(0)}(-l_1, -l_2, -l_3, 3) \, {\cal
A}_5^{\rm MHV,(0)}(1, 2, l_3, l_2, l_1) \Big] \, ,
\eea
where in the last equation we have used supersymmetric amplitudes and form factors \cite{harmony}
in order to perform the sum over internal helicities efficiently.
The fermionic integration can now be performed easily,  and after switching back
to component amplitudes and form factors the result is
\bea && {\cal F}_3^{\rm MHV,(0)}(1,2,3) \bigg[ {\langle 12 \rangle
\langle 23 \rangle \langle 31 \rangle \over \langle l_1 l_2 \rangle
\langle l_2 l_3 \rangle \langle l_3 3 \rangle \langle 3 l_1 \rangle}
\, A_5^{\rm NMHV,(0)}(1^+, 2^+, l_3^-, l_2^-, l_1^-) \nonumber\\
&& \qquad \qquad  \qquad + {\langle 23 \rangle \langle 31 \rangle
\over \langle 2 l_3 \rangle \langle l_3 l_2 \rangle \langle l_2 l_1
\rangle \langle l_1 1 \rangle} \, F_{4,{\rm SD}}^{\rm
NMHV,(0)}(-l_1^-, -l_2^-, -l_3^-, 3^+) \bigg] \, .\eea
Note that the amplitudes and form factor  in the bracket are the bosonic components with fixed
helicities. $F_{4,{\rm SD}}^{\rm NMHV,(0)}$ is the form factor with
an insertion of the  operator ${\rm Tr}(F_{SD}^2) + \ldots$ (the dots refer to additional cubic and quartic terms)  which is part of
the stress-energy multiplet. The appearance of this
form factor is due to the fact that both ${\cal F}_{4,{\rm SD}}^{\rm
NMHV,(0)}$ and ${\cal F}_{4}^{\rm NMHV,(0)}$ share the same NMHV
factor, i.e.
\be R_{4}^{\rm NMHV,(0)} = {{\cal F}_{4,{\rm SD}}^{\rm NMHV,(0)}
\over {\cal F}_{4,{\rm SD}}^{\rm MHV,(0)} } =  {{\cal F}_{4}^{\rm
NMHV,(0)} \over {\cal F}_{4}^{\rm MHV,(0)} }~,  \ee
where the former is the super form factor with an insertion of the
Lagrangian ${\rm Tr}(F_{SD}^2) + \cdots$ and the latter with
${\rm
Tr}(\phi^2)$  \cite{harmony}.
By performing the fermionic integrations we extract the
$\eta_{l_1}^4 \eta_{l_2}^4 \eta_{l_3}^4$ component,  and obtain
$F_{4,{\rm SD}}^{\rm NMHV,(0)}(-l_1^-, -l_2^-, -l_3^-, 3^+)$.
This form factor can be calculated using  BCFW recursion
relations as in \cite{harmony},  with the result
\bea F_{4,{\rm SD}}^{\rm NMHV,(0)}(1^-, 2^-, 3^-, 4^+) &=& {\langle
3|p_{12}|4]^3 \over s_{412} [41] [12] \langle 3|p_{41}|2]} +
{\langle 1|p_{23}|4]^3 \over s_{234} [23] [34] \langle 1|p_{34}|2]}
\\ && + {\langle 13 \rangle^4 s_{1234}^2 \over s_{341}
\langle 34 \rangle \langle 41 \rangle \langle 3|p_{41}|2] \langle
1|p_{34}|2] } ~. \nonumber \eea
%
%
\begin{center}
\begin{figure}[t]
\centerline{\includegraphics[height=10cm]{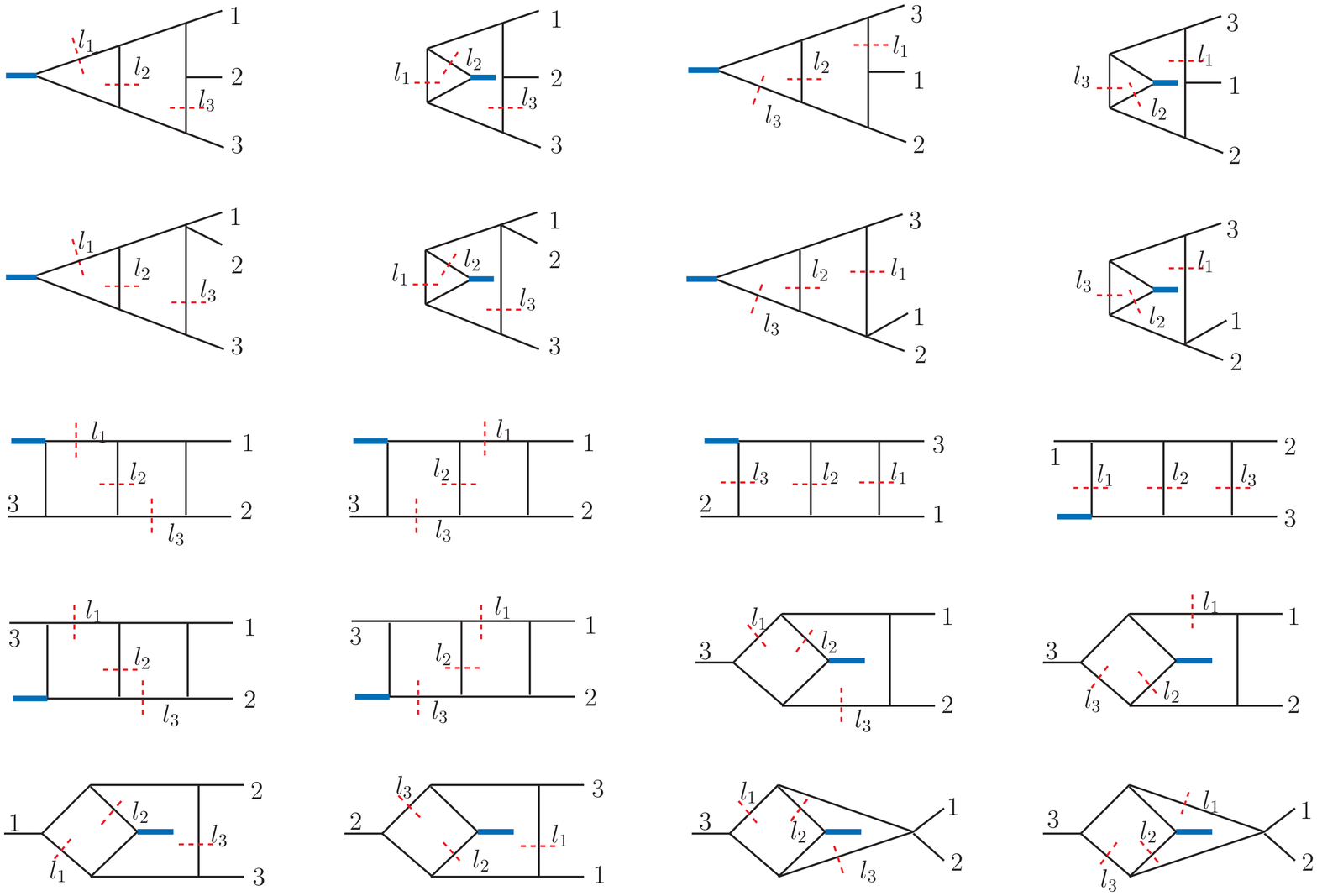} }
\caption{\it There are twenty integrals contributing to the
three-particle cut example discussed in the text.}
\label{3cutsexample}
\end{figure}
\end{center}
%
%
We compared this cut integrand with the expression obtained from the
three-particle cut of the integral presentation of the final result
given in  Figure \ref{fig1}. As shown in Figure \ref{3cutsexample},
there are in total 20 contributing cut integrals. Remarkably,  this
cut involves all integral topologies appearing in Figure \ref{fig1}
and we find complete agreement with the above non-trivial cut
integrand%
\footnote{Note that  no parity-odd terms can be present 
in the final result for the three-point form factor since  there are not enough independent momenta to 
form  a non-vanishing contraction with $\eps_{\mu \nu \rho \lambda}$. 
For the same
reason we also expect there should be no  $\mu^2$-terms in the expressions of the cuts,
and hence it is sufficient to use four-dimensional unitarity.}. This
provides one of the strongest consistency checks that our result
summarised in Figure \ref{fig1} is indeed correct.

\begin{center}
\begin{figure}[t]
\centerline{\includegraphics[height=3.5cm]{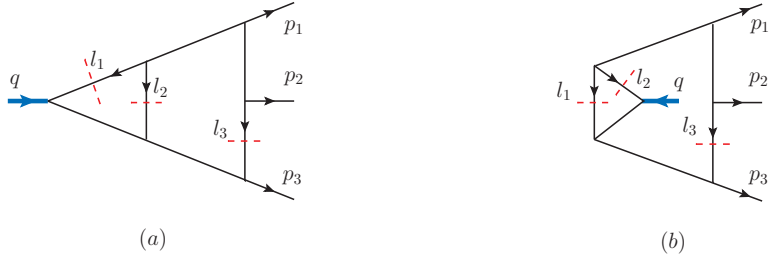} }
\caption{\it The two appearances of the integral TriPent in the
three-particle cut. They give different contributions  to the cut
integrand.} \label{TriPent2ways}
\end{figure}
\end{center}

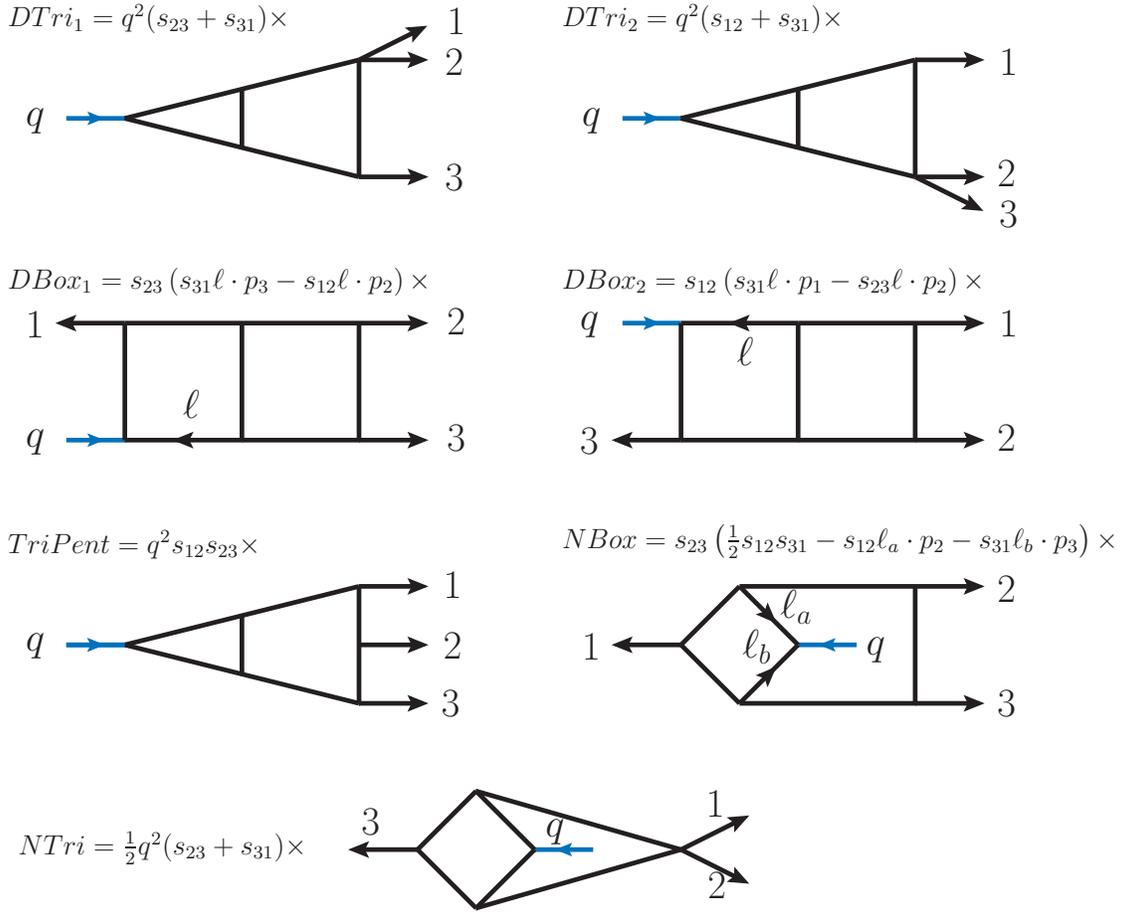
\begin{figure}[h]
\begin{center}
\scalebox{0.88}{ \centering \fcolorbox{white}{white}{
  \begin{picture}(451,394) (2,-46)
    \SetWidth{2.0}
    \SetColor{Black}
    \Line[arrow,arrowpos=0,arrowlength=6.667,arrowwidth=2.667,arrowinset=0.2,flip](150.425,-19.587)(175.496,-19.587)
    \Line(338.457,306.335)(338.457,281.264)
    \Line(150.425,318.87)(150.425,268.728)
    \Line(100.283,306.335)(100.283,281.264)
    \SetColor{RoyalBlue}
    \Line[arrow,arrowpos=0.5,arrowlength=5,arrowwidth=2,arrowinset=0.2](25.071,293.799)(50.142,293.799)
    \SetColor{Black}
    \Line(50.142,293.799)(150.425,318.87)
    \Line(50.142,293.799)(150.425,268.728)
    \Line[arrow,arrowpos=1,arrowlength=6.667,arrowwidth=2.667,arrowinset=0.2](150.425,318.87)(175.496,331.405)
    \Line[arrow,arrowpos=1,arrowlength=6.667,arrowwidth=2.667,arrowinset=0.2](150.425,318.87)(175.496,318.87)
    \Line[arrow,arrowpos=1,arrowlength=6.667,arrowwidth=2.667,arrowinset=0.2](150.425,268.728)(175.496,268.728)
    \Text(7.835,288.315)[lb]{\Large{\Black{$q$}}}
    \Text(188.815,331.405)[lb]{\Large{\Black{$1$}}}
    \Text(188.031,313.386)[lb]{\Large{\Black{$2$}}}
    \Text(188.031,264.028)[lb]{\Large{\Black{$3$}}}
    \SetColor{RoyalBlue}
    \Line[arrow,arrowpos=0.5,arrowlength=5,arrowwidth=2,arrowinset=0.2](263.244,293.799)(288.315,293.799)
    \SetColor{Black}
    \Line(288.315,293.799)(388.598,318.87)
    \Line(288.315,293.799)(388.598,268.728)
    \Line(388.598,318.87)(388.598,268.728)
    \Line[arrow,arrowpos=1,arrowlength=6.667,arrowwidth=2.667,arrowinset=0.2](388.598,318.87)(413.669,318.87)
    \Line[arrow,arrowpos=1,arrowlength=6.667,arrowwidth=2.667,arrowinset=0.2](388.598,268.728)(413.669,268.728)
    \Line[arrow,arrowpos=1,arrowlength=6.667,arrowwidth=2.667,arrowinset=0.2](388.598,268.728)(413.669,256.193)
    \Text(246.791,288.315)[lb]{\Large{\Black{$q$}}}
    \Text(426.205,313.386)[lb]{\Large{\Black{$1$}}}
    \Text(425.421,265.594)[lb]{\Large{\Black{$2$}}}
    \Text(426.205,248.358)[lb]{\Large{\Black{$3$}}}
    \SetColor{RoyalBlue}
    \Line[arrow,arrowpos=0.5,arrowlength=5,arrowwidth=2,arrowinset=0.2](25.071,155.909)(50.142,155.909)
    \SetColor{Black}
    \Line(50.142,155.909)(50.142,206.051)
    \Line[arrow,arrowpos=1,arrowlength=6.667,arrowwidth=2.667,arrowinset=0.2](50.142,206.051)(175.496,206.051)
    \Line[arrow,arrowpos=1,arrowlength=6.667,arrowwidth=2.667,arrowinset=0.2](50.142,206.051)(25.071,206.051)
    \Line(150.425,155.909)(150.425,206.051)
    \Line(100.283,155.909)(100.283,206.051)
    \Text(7.835,200.567)[lb]{\Large{\Black{$1$}}}
    \Text(188.815,202.134)[lb]{\Large{\Black{$2$}}}
    \Text(188.815,151.992)[lb]{\Large{\Black{$3$}}}
    \Text(7.835,150.425)[lb]{\Large{\Black{$q$}}}
    \SetColor{RoyalBlue}
    \Line[arrow,arrowpos=0.5,arrowlength=5,arrowwidth=2,arrowinset=0.2](263.244,206.051)(288.315,206.051)
    \SetColor{Black}
    \Line[arrow,arrowpos=0,arrowlength=6.667,arrowwidth=2.667,arrowinset=0.2,flip](263.244,155.909)(288.315,155.909)
    \Line[arrow,arrowpos=1,arrowlength=6.667,arrowwidth=2.667,arrowinset=0.2](288.315,155.909)(413.669,155.909)
    \Line(288.315,155.909)(288.315,206.051)
    \Line(388.598,155.909)(388.598,206.051)
    \Line(338.457,155.909)(338.457,206.051)
    \Text(246.008,200.567)[lb]{\Large{\Black{$q$}}}
    \Text(426.205,201.35)[lb]{\Large{\Black{$1$}}}
    \Text(425.421,151.992)[lb]{\Large{\Black{$2$}}}
    \Text(246.008,150.425)[lb]{\Large{\Black{$3$}}}
    \SetColor{RoyalBlue}
    \Line[arrow,arrowpos=0.5,arrowlength=5,arrowwidth=2,arrowinset=0.2](25.071,68.161)(50.142,68.161)
    \SetColor{Black}
    \Line[arrow,arrowpos=1,arrowlength=6.667,arrowwidth=2.667,arrowinset=0.2](150.425,93.232)(175.496,93.232)
    \Line[arrow,arrowpos=1,arrowlength=6.667,arrowwidth=2.667,arrowinset=0.2](150.425,68.161)(175.496,68.161)
    \Line[arrow,arrowpos=1,arrowlength=6.667,arrowwidth=2.667,arrowinset=0.2](150.425,43.091)(175.496,43.091)
    \Line(150.425,43.091)(150.425,93.232)
    \Line(50.142,68.161)(150.425,93.232)
    \Line(50.142,68.161)(150.425,43.091)
    \Line(100.283,55.626)(100.283,80.697)
    \Text(7.835,62.677)[lb]{\Large{\Black{$q$}}}
    \Text(187.248,88.531)[lb]{\Large{\Black{$1$}}}
    \Text(187.248,62.677)[lb]{\Large{\Black{$2$}}}
    \Text(186.465,37.606)[lb]{\Large{\Black{$3$}}}
    \SetColor{RoyalBlue}
    \Line[arrow,arrowpos=0.5,arrowlength=5,arrowwidth=2,arrowinset=0.2,flip](338.457,68.161)(363.528,68.161)
    \SetColor{Black}
    \Line(288.315,68.161)(313.386,93.232)
    \Line(288.315,68.161)(313.386,43.091)
    \Line[arrow,arrowpos=1,arrowlength=6.667,arrowwidth=2.667,arrowinset=0.2](313.386,93.232)(413.669,93.232)
    \Line[arrow,arrowpos=1,arrowlength=6.667,arrowwidth=2.667,arrowinset=0.2](313.386,43.091)(413.669,43.091)
    \Line(388.598,43.091)(388.598,93.232)
    \Line[arrow,arrowpos=0.5,arrowlength=6.667,arrowwidth=2.667,arrowinset=0.2](313.386,93.232)(338.457,68.161)
    \Line[arrow,arrowpos=0.5,arrowlength=6.667,arrowwidth=2.667,arrowinset=0.2](313.386,43.091)(338.457,68.161)
    \Line[arrow,arrowpos=1,arrowlength=6.667,arrowwidth=2.667,arrowinset=0.2](288.315,68.161)(263.244,68.161)
    \Text(246.791,61.894)[lb]{\Large{\Black{$1$}}}
    \Text(425.421,86.965)[lb]{\Large{\Black{$2$}}}
    \Text(425.421,37.606)[lb]{\Large{\Black{$3$}}}
    \Text(369.012,61.11)[lb]{\Large{\Black{$q$}}}
    \Text(0,331.405)[lb]{{\Black{$DTri_1 = q^2(s_{23}+s_{31}) \times$}}}
    \Text(238.173,331.405)[lb]{{\Black{$DTri_2 =  q^2 (s_{12}+s_{31}) \times$}}}
    \Text(0,218.587)[lb]{{\Black{$DBox_1=s_{23} \left( s_{31} \ell \cdot p_3 - s_{12} \ell \cdot p_2\right) \times$}}}
    \Text(238.173,218.587)[lb]{{\Black{$DBox_2=s_{12} \left( s_{31} \ell \cdot p_1-s_{23} \ell \cdot p_2 \right) \times$}}}
    \Text(0,105.768)[lb]{{\Black{$TriPent = q^2 s_{12} s_{23} \times$}}}
    \Text(238.173,105.768)[lb]{{\Black{$NBox=s_{23}\left({1\over 2} s_{12} s_{31} - s_{12} \ell_a \cdot p_2 - s_{31} \ell_b \cdot p_3\right) \times$}}}
    \Line[arrow,arrowpos=0.5,arrowlength=6.667,arrowwidth=2.667,arrowinset=0.2,flip](50.142,155.909)(100.283,155.909)
    \Line[arrow,arrowpos=1,arrowlength=6.667,arrowwidth=2.667,arrowinset=0.2](100.283,155.909)(175.496,155.909)
    \Text(75.213,166.094)[lb]{\Large{\Black{$\ell$}}}
    \Line[arrow,arrowpos=1,arrowlength=6.667,arrowwidth=2.667,arrowinset=0.2](338.457,206.051)(413.669,206.051)
    \Line[arrow,arrowpos=0.5,arrowlength=6.667,arrowwidth=2.667,arrowinset=0.2](338.457,206.051)(288.315,206.051)
    \Text(313.386,188.815)[lb]{\Large{\Black{$\ell$}}}
    \Text(332.189,77.563)[lb]{\Large{\Black{$\ell_a$}}}
    \Text(315.736,60.327)[lb]{\Large{\Black{$\ell_b$}}}
    \SetColor{RoyalBlue}
    \Line[arrow,arrowpos=0.5,arrowlength=5,arrowwidth=2,arrowinset=0.2](25.071,68.161)(50.142,68.161)
    \Line[arrow,arrowpos=0.5,arrowlength=5,arrowwidth=2,arrowinset=0.2,flip](225.638,-19.587)(250.709,-19.587)
    \SetColor{Black}
    \Line(175.496,-19.587)(200.567,5.484)
    \Line(175.496,-19.587)(200.567,-44.657)
    \Line(200.567,-44.657)(288.315,-19.587)
    \Line(200.567,5.484)(288.315,-19.587)
    \Line[arrow,arrowpos=1,arrowlength=6.667,arrowwidth=2.667,arrowinset=0.2](288.315,-19.587)(313.386,-7.051)
    \Line[arrow,arrowpos=1,arrowlength=6.667,arrowwidth=2.667,arrowinset=0.2](288.315,-19.587)(313.386,-32.122)
    \Text(300.067,-5.484)[lb]{\Large{\Black{$1$}}}
    \Text(300.85,-40.74)[lb]{\Large{\Black{$2$}}}
    \Text(151.992,-13.319)[lb]{\Large{\Black{$3$}}}
    \Text(4.701,-25.854)[lb]{{\Black{$NTri=\frac{1}{2}q^2 (s_{23}+s_{31}) \times$}}}
    \Line(200.567,-44.657)(225.638,-19.587)
    \Line(200.567,5.484)(225.638,-19.587)
    \Text(231.122,-17.236)[lb]{\Large{\Black{$q$}}}
  \end{picture}
}
}
\end{center}
\caption[]%
  {\it The integral expansion of our final result for the three-point form factor ${\mathcal G}^{(2)}_{3}$.}
\label{fig1}
\end{figure}

There is an  important subtlety we would like  to point out.
As we mentioned before, we used double two-particle cuts  on
(\ref{ffc3p}) in order to first detect all the integrals which can
appear. We find that $TriPent$ and $DTri$ actually receive two
identical contributions from  the single-trace and the double-trace
term in (\ref{ffc3p}). We have drawn the two contributions from the
planar and non-planar part in Figures \ref{TriPent2ways}$(a)$ and
\ref{TriPent2ways}$(b)$, respectively. Similar comments apply to
$DTri$. This observation is relevant  when we consider unitarity
cuts. For example, in the three-particle example we considered
above, two different presentation of $TriPent$ give a  different
contribution to the cut integrand -- indeed,  as shown in Figure
\ref{TriPent2ways}, the positions of $l_1$ and $l_2$ are swapped,
and therefore the total contribution to the cut integrand is
\be { 1\over2} {{ q^2 s_{12} s_{23} } \over { (l_3+p_2)^2
(l_3-p_3)^2 (l_1+l_2)^2 (l_1+q)^2}} +  { 1\over2} {{ q^2 s_{12}
s_{23} } \over { (l_3+p_2)^2 (l_3-p_3)^2 (l_1+l_2)^2 (l_2+q)^2}} ~.
\ee
%
%
%
%

We conclude this section by presenting our complete result for the
two-loop three-particle form factor in $\cN =4$ SYM obtained from
generalised unitarity cuts: \be\label{result} {\mathcal
G}^{(2)}_{3}=\sum_{i=1}^2 (DTri_i+DBox_i)+TriPent+NBox+NTri +
\mathrm{cyclic} \, , \ee
where in Figure \ref{fig1} we have presented  compactly all
integrals, including their precise coefficients and numerators  some
of which are loop dependent.

The integral functions appearing in \eqref{result} have six or seven
propagators and notably both planar and non-planar topologies
appear, some with irreducible, loop-dependent numerators. The
maximum power of loop momentum appearing in the numerators is one.
We will now proceed to discuss the numerical evaluation of these
integrals.



\subsection{Evaluation of the integral functions}
\label{3.1}

Several  of the integral functions appearing in our  result \eqref{result}  have been computed  and explicit expressions can be found in \cite{gr99, grp, grnp} -- in particular all the six-propagator integrals are known.
For the remaining integrals $TriPent, DBox_i$ and $NBox$ explicit analytic expressions are not known, in part due to the peculiar numerators we found. In principle, there exist algorithms that allow to reduce these integrals to known master integrals  \cite{gr99,grp,grnp}.
However, in this work we chose to use an independent route, and expressed them in terms of Mellin-Barnes (MB) representations (see \cite{smirnov} for a pedagogical introduction) and used well-established codes  \cite{czakon} to evaluate them numerically to high precision.

The MB form of the planar integrals can readily be constructed with public programmes like AMBRE  \cite{ambre1,ambre2} and we will not quote their explicit forms here. However, the MB representation of the $NBox$ integral cannot be found with AMBRE due to well-known issues of this programme with non-planar topologies with several scales.
Therefore, we constructed an  MB representation of $NBox$ directly from its Feynman parameter form. The result is an eight-fold MB representation of the form
\bea\label{npbox}
&&\frac{(-q^2)^{-2 \eps}}{2 (2\pi i)^8 \Gamma(-1 - 3 \eps) } \int \prod_{i=1}^8 \left(dz_i \Gamma(-z_i)\right)
u^{z_5+1} v^{z_{678}+1}
     w^{-3 - 2 \eps - z_{12345678}} \times \nonumber \\
&&   \Gamma(-\eps - z_{34}) \Gamma(-\eps + z_4) \Gamma(1 + z_{13456})
\Gamma(1 + z_{157}) \Gamma(-1 - \eps + z_3 - z_8)  \times \nonumber \\
&& \Gamma(-2 - 2 \eps -
      z_1 - z_{568}) \Gamma(-2 - 2 \eps - z_{134578})
      \Gamma(-2 - 2 \eps - z_{1234678}) \times \\
&&      \frac{\Gamma(-2 \eps - z_3 + z_8) \Gamma(1 + z_{168}) \Gamma(1 + z_{278})
\Gamma(3 + 2 \eps + z_{12345678})}{\Gamma(-2 \eps - z_3)
\Gamma(-1 - 2 \eps -
      z_3 - z_{48}) \Gamma(-1 - 2 \eps + z_{34} - z_8) \Gamma(-2 \eps -
      z_{34} + z_8)} \, ,
      \nonumber
\eea
where we have introduced the shorthand notation $z_{ij \dots k} = z_i+z_j +\ldots + z_k$,  and
\beq
\label{uvw}
u\, =:\, {s_{12}\over q^2}\, , \qquad
v\, :=\, {s_{23}\over q^2}\,, \qquad w\, :=\, {s_{31}\over q^2}\, .
\eeq
Note that for sake of brevity we have dropped here the terms of the numerator which are linear in loop momentum $\ell$; they lead to a number of similar eight-fold MB integrals. Furthermore, due to the $\Gamma(-1 - 3 \eps)$ denominator the integral effectively becomes seven-dimensional \cite{smirnov}. In this sense this integral is the most complicated and numerically the most challenging
contribution to the form factor,  since the planar topologies $DBox_i$ and $TriPent$ require at most three- and four-fold MB integrals, respectively.

We have evaluated \eqref{result} by expressing all six-propagator integrals by their
known analytic formulae and using {\tt MB.m} by \cite{czakon} for numerical evaluations
of the MB representations of all seven-propagator topologies. The result is a power series
starting as $\eps^{-4}$ which we have computed up to and including finite ${\mathcal O}(\eps^0)$ terms. We present here a few  results of our numerical evaluation at four kinematic points $(-s_{12},-s_{23},-s_{31})$:
\bea
(1,1,1): && \frac{4.5}{\eps^4}+\frac{0.}{\eps^3}+\frac{6.12223}{\eps^2}-\frac{16.7052}{\eps}-18.2484 \pm 0.02 + \mathcal{O}(\eps)\, ,  \\
(1,1,2): && \frac{4.5}{\eps^4}-\frac{2.07944}{\eps^3}+\frac{7.98765}{\eps^2}-\frac{18.9491}{\eps}-7.3182 \pm 0.02 +\mathcal{O}(\eps)\, ,  \nonumber \\
(1,2,2): &&
   \frac{4.5}{\eps^4}-\frac{4.15888}{\eps^3}+\frac{9.2099}{\eps^2}-\frac{23.0025}{\eps}+ 1.8686 \pm 0.02 + \mathcal{O}(\eps)\, ,
\nonumber \\
(1,2,3): &&
   \frac{4.5}{\eps^4}-\frac{5.37528}{\eps^3}+\frac{11.6703}{\eps^2}-\frac{25.9714}{\eps}+ 10.6624\, \pm 0.03 +\mathcal{O}(\eps)\, . \nonumber
\eea
Here we have only quoted the errors of the finite terms. The numerical error of the $1/\eps^4$ and $1/\eps^3$ are negligible, while for all kinematic points we have investigated, a conservative estimate of the errors of the $1/\eps^2$ and $1/\eps$ terms is
$10^{-13}$  and $10^{-7}$,  respectively.

In the following we will study this two-loop result in detail and, motivated by the fact that form factors have universal collinear limits,
we will define  a finite remainder very much in the spirit of \cite{abdk,bds}.  We will then determine the symbol  \cite{symbol} of the remainder function, and from that derive its analytic expression, which we will compare to our numerical results.


\section{Exponentiation of the form factor and the remainder function}

In this section we want to consider the possibility that higher-loop form factors in $\cN=4$ SYM obey a similar
exponentiation relation as MHV loop amplitudes \cite{abdk,bds} for small numbers of on-shell particles,  and understand
from which number of particles a remainder function has to be added.
We expect this remainder to appear for a smaller number of particles compared to amplitudes because only Lorentz symmetry and dilatations are unbroken (up to infrared divergences). This leaves us with $3n-7$ parameters on which a remainder might depend. Indeed we will find that the three-point remainder function depends on two variables. Similar observations were made at strong coupling for $(1+1)$-dimensional kinematics in \cite{mz}, where  a non-trivial remainder appears at four points.

For the Sudakov form factor, exponentiation is evidently true even in QCD, and in $\cN=4$ this was explicitly proved in \cite{VN1}.
Specifically, one finds
\beq
\label{orig}
 F^{(2)} (q^2, \eps) - {1\over 2} \big( F^{(1)} (q^2, \eps)\big)^2 \ = \ (-q^2)^{- 2 \eps} \left[
 {\zeta_2 \over \eps^2} + {\zeta_3 \over \eps} + \cO (\eps)\right] \ ,
 \eeq
where $F^{(L)}$ is the $L$-loop Sudakov form factor.
The result in \eqref{orig}  was re-derived in appendix B of \cite{bsty}, where it was recast  in a slightly more modern language as
 \beq
 \label{modern}
 F^{(2)} (q^2, \eps) - {1\over 2} \big( F^{(1)} (q^2, \eps)\big)^2 \ = \  f^{(2)} (\eps) F^{(1)} (q^2, 2 \eps) +  C^{(2)} +  \cO (\eps)\ ,
 \eeq
where $f^{(2)}  (\eps) = f_0^{(2)} + f_1^{(2)} \eps + f_2^{(2)} \eps^2$, with
\beq
f_0^{(2)} = - 2 \zeta_2\ , \qquad f_1^{(2)} = - 2 \zeta_3 \ ,
\eeq
and with one relation between $f_2^{(2)}$ and $C^{(2)} $, namely%
\footnote{We note that $f_2^{(2)}$ will be fixed later in \eqref{f2fixedaswell} from collinear factorisation.}
\beq
\label{onerel}
C^{(2)}   = { f_2^{(2)}\over 2} +  {\pi^4\over 18}
\ .
\eeq
For comparison with amplitudes, we recall that the four-point MHV amplitude (divided by the tree-level amplitude) satisfies \cite{abdk, bds}
 \beq
 \label{su}
\cM^{(2)}_4  ( \eps) - {1\over 2} \big( \cM^{(1)}_4 ( \eps)\big)^2 \ = \  f^{(2)}_{\rm amp}  (\eps) \cM^{(1)}_4 (2 \eps) +  C^{(2)}_{\rm amp} +  \cO (\eps)\ ,
 \eeq
with $f^{(2)}_{\rm amp}  (\eps)= \tilde{f}_0^{(2)} + \eps \tilde{f}_1^{(2)} +\eps^2 \tilde{f}_2^{(2)}$,  and%
\footnote{As observed in \cite{bsty}, the factor of 1/2 in the  result \eqref{dd2} is a matter of convention -- it can be understood once one recalls that $f_0^{(2)}$ and $f_1^{(2)}$ are written in a convention where the 't Hooft coupling $a_{\rm BDS}$ is twice as that used in  \cite{VN1, bsty}. It is the combination  $a f^{(2)}$ which must be independent of any conventions used to define the coupling.}
\beq
\label{dd2}
\tilde{f}_0^{(2)} = -\zeta_2 =  {f_0^{(2)} \over 2} \ , \qquad \tilde{f}_1^{(2)} = - \zeta_{3} = {f_1^{(2)} \over 2} \ .
\eeq

\subsection{Collinear factorisation}
In this section we would like to discuss collinear factorisation and the exponentiation of infrared divergences of form factors. This will lead  us  to the definition of  a finite, scaling-invariant and regulator-independent remainder function with trivial collinear limits.

We begin by recalling that an important hint that  MHV amplitudes in $\cN=4$ may exponentiate came from the study of collinear limits \cite{abdk} where two adjacent momenta $a$ and $b$ become parallel. It is well known that scattering amplitudes have a universal collinear factorisation behaviour, which is governed by splitting amplitudes \cite{bddk, fusing,Bern:1995ix}.
 These quantities only depend on the helicities and the momentum fractions carried by the two legs becoming collinear, but are  completely blind to other details of the process. If we consider $\cN=4$ SYM and focus on the case of MHV amplitudes,
the helicity-blind ratio $A^{(L)}_n/A^{(0)}_n = \cM^{(L)}_n(p_1,  \ldots ,  p_n)$ therefore obeys the following factorisation:
\be
\cM^{(L)}_n(p_1,  \ldots p_a, p_b,  \ldots ,  p_n)
{\buildrel a \parallel b\over
{\relbar\mskip-1mu\joinrel\longrightarrow}}
 \sum_{l=0}^L \cM^{(l)}_{n-1}(p_1,  \ldots,  p_a+p_b , \ldots ,  p_n) r^{(L-l)}(\e; z, s_{ab}) \ ,
 \label{collfact}
\ee
where $r^{(0)} = 1$, $\cM_n^{(0)}=1$, and
\be
\label{sopraa}
r^{(1)}(\e; z, s_{ab}) \, :=\,
{c_\Gamma \over \e^2} \Big( {-s_{ab} \over \mu^2} \Big)^{-\e}
\left[  1 \, - \,
\mbox{}_{2}F_1 \left( 1, -\e, 1- \e, {z-1 \over z}\right)
  \, - \, \mbox{}_{2}F_1 \left(1, -\e, 1- \e, {z \over z-1}\right)
\right] \, ,
\ee
to all orders in the dimensional regularisation parameter $\e$ \cite{ku,vittorio}.%
\footnote{The all-orders in $\eps$ expression for $r^{(1)}(\e; z, s_{ab})$ in \eqref{sopraa} was also rederived in \cite{ftt} using one-loop MHV diagrams.}

For general large-$N$ gauge theories, a unitarity-based proof   of \eqref{collfact} at any loop order has been given in \cite{collproof}. The important point we would like to make is that  the main steps of that proof apply directly to form factors as well,  which therefore share the universal factorisation properties of amplitudes, including  \eqref{collfact}, which we write as
\be
\cG^{(L)}_n(p_1,  \ldots p_a, p_b,  \ldots ,  p_n)
{\buildrel a \parallel b\over
{\relbar\mskip-1mu\joinrel\longrightarrow}}
 \sum_{l=0}^L \cG^{(l)}_{n-1}(p_1 , \ldots , p_a+p_b , \ldots , p_n) r^{(L-l)}(\e; z, s_{ab}) \ ,
 \label{collfactFF}
\ee
where we have defined
\beq
\label{ratioF}
\cG^{(L)}_n \ := \ F^{(L)}_n/F^{(0)}_n
\ .
\eeq
We stress that the splitting amplitudes appearing  in \eqref{collfactFF} are the same as for amplitudes.%
\footnote{We have checked this for bilinear operators following the proof of \cite{collproof}, however  we caution the reader that the proof presented in that paper  does not extend automatically to operators such as  $\Tr \phi^n$ for $n>2$, and collinear factorisation should be re-examined carefully in these cases. }

Remarkably, loop splitting amplitudes in $\cN=4$ SYM obey a cross-order relation which at two loops is \cite{abdk}
\be
\label{babis2}
r^{(2)}(\e; z, s)\ = \ \frac{1}{2} \left( r^{(1)}(\e; z, s)\right)^2 \, + \, f^{(2)} (\e) r^{(1)}(2\e; z, s)\, + \, \cO (\eps)
\ ,
\ee
which has generalisations to higher loops \cite{bds}. Importantly, this implies that the correct collinear factorisation of the two-loop amplitude $\cM_n^{[2]}$ is captured by the combination
\be
\widetilde{\cM}_n^{(2)}\ :=\ \frac{1}{2} (\cM_n^{(1)}(\e))^2 \, + \, f^{(2)}(\e) \cM_n^{(1)} (2 \eps)
\ ,
\ee
which is given purely in terms of the one-loop ratio functions and $f^{(2)} (\e)$. The only allowed discrepancy between $\widetilde{\cM}_n^{(2)}$ and the full two-loop ratio $\cM_n^{(2)}$ would be
a function that is finite in all collinear limits.  As is well known, in $\cN=4$ SYM this difference turns out to be a constant for $n=4,5$ and a non-trivial function $n>5$  -- the remainder function.
This behaviour is linked to the fact that loop amplitudes in $\cN=4$ SYM obey an anomalous dual conformal Ward identity \cite{dhks-conf, dhks}, of which
$\widetilde{\cM}_n^{(2)}$  is a particular solution. For $n=4, 5$ no dual conformal cross-ratios exists, and the only homogeneous solution  is a constant, while for $n>5$ there are $3 n - 15$ cross ratios and explicit calculations have confirmed the necessity to add a remainder function ${\mathcal R}_n(u_1,  \ldots ,  u_{3n-15})$ \cite{seven, dhkshex}.

Given the similarities between amplitudes and form factors, specifically concerning their collinear factorisation properties and the universality and exponentiation of their  infrared divergences \cite{magician, Magnea:1990zb,Sterman:2002qn,Becher:2009cu,Gardi:2009qi}, it is natural to expect that loop form factors should have a similar exponentiated form. Let us investigate this point in more detail.

To begin with, we have checked explicitly in a specific example our claim based on a simple generalisation of \cite{collproof} that form factors enjoy collinear factorisation properties identical to those of amplitudes. Specifically, we have confirmed that  the one-loop MHV form factors derived in \cite{bsty} indeed obey the anticipated  collinear factorisation \eqref{collfactFF}.
For concreteness we now  consider the one-loop three-point form factor ratio function (see (3.17) in \cite{bsty}),
\bea
 \label{3pt1loop}
\cG^{(1)}_3 & = & -\frac{c_\Gamma}{\e^2} \Big[(-s_{12})^{-\e}+
(-s_{23})^{-\e}+ (-s_{31})^{-\e}  \\
 & & + (-s_{12})^{-\e} h\Big(-\frac{s_{31}}{s_{23}}\Big) + (-s_{23})^{-\e} h\Big(-\frac{s_{31}}{s_{12}}\Big) -
 (-q^2)^{-\e} h\Big(-\frac{s_{31} q^2}{s_{12} s_{23}}\Big) \nonumber \\
 & & + (-s_{23})^{-\e} h\Big(-\frac{s_{12}}{s_{31}}\Big) + (-s_{31})^{-\e} h\Big(-\frac{s_{12}}{s_{23}}\Big) -
 (-q^2)^{-\e} h(-\frac{s_{12} q^2}{s_{23} s_{31}}) \nonumber \\
 & & + (-s_{31})^{-\e} h\Big(-\frac{s_{23}}{s_{12}}\Big) + (-s_{12})^{-\e} h\Big(-\frac{s_{23}}{s_{31}}\Big) -
 (-q^2)^{-\e} h\Big(-\frac{s_{23} q^2}{s_{31} s_{12}}\Big) \Big]\, ,
 \nonumber
\eea
with $h(x) = \mbox{}_2F_1(1,-\e,1-\e,x)-1$ and $q^2 = (p_1+p_2+p_3)^2 = s_{12}+s_{23}+s_{31}$.

In the collinear limit $1 || 2$ we have $s_{12} \to 0$, $q^2 \to s_{23}+s_{31}$. Setting, in the limit, $s_{31}/q^2 \to z$ and $s_{23}/q^2 \to 1-z$,  we find
\be
\cG^{(1)}_3\
{\buildrel 1 \parallel 2 \over
{\relbar\mskip-1mu\joinrel\longrightarrow}}\
r^{(1)}(\e; s_{12},z) \, - \, \frac{2 c_\Gamma}{\e^2} (-(P+p_3)^2)^{-\e} \ ,
\ee
with $P=p_1+p_2$. Note that the splitting amplitude comes from terms 1, 4 and 11 in \eqref{3pt1loop}, and the second term is nothing but the one-loop two-point (Sudakov) form factor. Reproducing the expected factorisation requires a delicate conspiracy among  various terms in \eqref{3pt1loop}.


\subsection{Iterative structure at higher loops}

Having established the correct collinear factorisation at one loop,  we can now write down
a general two-loop MHV form factor ansatz as follows,
\be
\label{fm}
\cG^{(2)}_n \  = \ \frac{1}{2} \big(\cG^{(1)}_n(\e)\big)^2 \, + \, f^{(2)} (\e) \, \cG^{(1)}_n(2 \e)  \, + \, {\mathcal R}_n^{(2)}\, ,
\ee
where the ratio $\cG^{(L)}_n$ is defined in \eqref{ratioF}. Note that we allow for a potential form factor remainder function ${\mathcal R}_n^{(2)}$ on the right-hand side of \eqref{fm}.
 A few key properties of the form factor remainder are:

\begin{itemize}
\item[{\bf 1.}] It must free of infrared divergences,  and

\item[{\bf 2.}]
It must be finite in all collinear limits. Upon properly normalising the remainder function by adding to it an $n$-independent, transcendentality-four constant, we expect that the  normalised remainder behaves as
\beq
\label{abbc}
  \cR_n \rightarrow \cR_{n-1} \ ,
\eeq
 in a simple collinear limit. Furthermore, we also expect that
\item[{\bf 3.}]
It must be expressed in terms of  transcendentality-four functions, of which we will shortly discuss the symbol.
\item[{\bf 4.}] It is a rescaling invariant function and hence it depends on Mandelstam variables only through their ratios. But due to the lack of dual conformal (super) symmetry of form factors it does not enjoy the more restricted dependence on conformal cross ratios of amplitude remainders in planar $\cN=4$ SYM. For this reason,  a non-trivial form factor remainder can already appear at $n=3$.

\end{itemize}

The first property follows from the exponentiation of infrared divergences. Pleasingly, we have checked it explicitly in the case of the three-point, two-loop form factor.

The second property follows from the analysis of collinear limits at higher loops \cite{seven}. This analysis was performed in the case of amplitudes in Section 7 of \cite{abhkst}, following \cite{seven}, and it is instantly extended to form factors because their collinear factorisation, as expressed in \eqref{collfactFF}, works in the same way as in the amplitude case \eqref{collfact}. In particular, using \eqref{collfactFF}, one  finds that under a simple collinear limit,  the scalar functions $\cG^{(1)}_n$  and $\cG^{(2)}_n$   behave as
\beqa
\label{prima}
\cG^{(1)}_n & \to  &  \cG^{(1)}_{n-1} + r^{(1)} (\e; z, s_{ab}) \ ,
\\ \nonumber
\cG^{(2)}_n & \to  &  \cG^{(2)}_{n-1}  + r^{(1)} (\e; z, s_{ab} ) \cG^{(1)}_{n-1} + r^{(2)} (\e; z, s_{ab} )
\ .
\eeqa
Using the fact  that splitting amplitudes obey an iterative formula identical to the
homogeneous form of the
BDS conjecture for the amplitude \cite{abdk},  expressed by  \eqref{babis2},
 we conclude that under a simple collinear limit,
\begin{align}
\label{vh}
\cG^{(2)}_n(\e ) &- {1\over 2} \big( \cG^{(1)}_n (\e) \big)^2 - f^{(2)}(\e)  \cG_n^{(1)} ( 2 \e ) \nonumber \\
 &\rightarrow\
\cG^{(2)}_{n-1}(\e ) - {1\over 2} \big( \cG^{(1)}_{n-1} (\e) \big)^2 - f^{(2)}(\e)  \cG_{n-1}^{(1)} ( 2 \e )
\ .
\end{align}
Equation \eqref{fm}
defines the form factor remainder  $\cR_n^{(2)}$ as
\beq
\label{fm2bis}
\cR_n^{(2)} \ := \    \cG^{(2)}_n(\e )\, - \, {1\over 2} \big( \cG^{(1)}_n (\e) \big)^2 -  f^{(2)} (\e) \cG^{(1)}_n  ( 2 \e ) \, - \,  C^{(2)}
\, + \cO (\e ) \
\ ,
\eeq
and it follows from  \eqref{vh} that in the simple collinear limit $\cR_n \   \to \   \cR_{n-1}$, as anticipated in
\eqref{abbc}. The outcome of this brief analysis is that writing the $L$-loop form factor as an $L$-loop BDS-like ansatz plus an $L$-loop remainder gives a very efficient way to organise the presentation of the result -- the BDS part captures infrared divergences and simple collinear factorisation, while the remainder, is finite in four dimensions and in the collinear limit flows smoothly from the $n$-point to the $n-1$-point remainder.
An important comment is in order here. The iterative structure in \eqref{vh} is driven by the iteration of the splitting amplitude \eqref{babis2}. For this reason, the function $f^{(2)}(\eps)$ is completely determined to be the same as in the amplitude case. In our normalisations for the coupling constant, we have
\beq
\label{f2fixedaswell}
f^{(2)} (\eps) \ = \ -2 \zeta_2 \, - \, 2 \zeta_3 \eps \, - \, 2 \zeta_4 \eps^2\ .
\eeq
In turn, this implies that we can determine the constant $C^{(2)}$ introduced in the iteration of the Sudakov form factor, see \eqref{modern}. Plugging  $f_2^{(2)} = - 2 \zeta_4$ into \eqref{onerel}, we find
\beq
C^{(2)} \ = \ 4\, \zeta_4
\ .
\eeq
Importantly, the constant $C^{(2)}$ is independent of the number of legs.
The consequence of this observation is that the three-point form factor remainder function as defined in \eqref{fm2bis} (for $n=3$) with the value for $C^{(2)}$ just found, must have a very precise collinear limit:
\beq
\cR^{(2)}_3 \ \rightarrow \ 0
\ ,
\eeq
in any of the three possible simple collinear limits.
The third property is simply a consequence of maximal transcendentality.

\subsection{Numerical results for the three-point remainder at two loops}

As explained in the previous section, it is natural to define a remainder function,  which at three points is given by
\beq
\label{G3}
\cR_3^{(2)} \ := \    \cG^{(2)}_3(\e )\, - \, {1\over 2} \big( \cG^{(1)}_3 (\e) \big)^2 \, - \,  f^{(2)} (\e)\,  \cG^{(1)}_3  ( 2 \e ) \, - \,  C^{(2)}
\, + \cO (\e ) \
\ .
\eeq
This quantity is finite, regulator-independent and vanishes in all soft and collinear limits.

Our  numerical evaluations confirmed that  all infrared-divergent $\eps^{-p}$ terms  in $\cR_3^{(2)}$ do indeed cancel: for $p=3,4$ numerical errors are negligible, while for $p = 2$ and $p = 1$ we have confirmed this up to about $10^{-13}$ and $10^{-7}$, respectively.
This is a stringent check that our result \eqref{result} obtained using generalised unitarity is indeed correct.
We have collected in Table 1 a few values of the remainder function obtained with our numerical programmes. The remainder is defined in \eqref{G3} where we have set $C^{(2)}=4 \zeta_4$ and $f^{(2)}_2=-2 \zeta_4$. We will later compare this numerical remainder to our analytical expression, see Table 2.

\begin{table}[h]
\begin{center}
\begin{tabular}{|c|c|c|}
\hline
$(u,v,w)$ & numerical $\cR_3^{(2)}$ & est. error \\
\hline \hline
(1/3, 1/3, 1/3) & -0.1519   & 0.02 \\
(1/4, 1/4, 1/2) & -0.1203   & 0.02 \\
(1/5, 2/5, 2/5) & -0.1301   & 0.02 \\
(1/2, 1/3, 1/6) & -0.1080   & 0.03 \\
\hline
\end{tabular}
\caption{\it Some numerical values of the remainder $\cR_3^{(2)}$ defined in \eqref{G3}, with $C^{(2)}=4 \zeta_4$ and $f^{(2)}=-2 \zeta_2 - 2  \zeta_3\eps  -2 \zeta_4\eps^2$. Infrared poles  cancel with negligible numerical errors. }
\end{center}
\end{table}


\subsection{The symbol of the three-point remainder at two loops}

The question we would like to discuss now is how we can constrain   the  form factor remainder function  ${\mathcal R}_n^{(2)}$ defined in \eqref{fm2bis} using its symbol \cite{symbol}.%
\footnote{Similar discussions have appeared for the symbol of amplitude remainder functions in
\cite{pulling, CaronHuot:2011ky, Dixon:2011pw, Heslop:2011hv, Dixon:2011nj, Prygarin:2011gd}.}
We will focus on the case of the three-point form factor calculated explicitly in this paper. The conclusion of the analysis presented in this section is that, in the three-point case,  there is a unique non-trivial solution for the symbol of the remainder function (up to an overall constant) to be presented shortly.

In order to write down an ansatz for the symbol of the remainder function, we make the following physical considerations which lead to important constraints on its structure:
\begin{itemize}
\item[{\bf 1.}]
We should understand which variables the remainder function can depend upon.
\item[{\bf 2.}]
The symbol must satisfy a certain first-entry condition \cite{pulling}, arising from the requirement that  the  two-loop form factor remainder must have  discontinuities only across the physical branch cuts starting at
$P_{J}^2 = 0$, where $P_{J}^2$'s are appropriate  kinematic invariants  made of sums of external momenta (in the three-point case considered here, these could be $s_{12}$, $s_{23}$ and $s_{13}$).
This implies that the symbol $\cS^{(2)}$ of the two-loop remainder must be of the form
\beq
\cS^{(2)} \  = \ \sum_{\rm disc_{\it J} } P_{J}^2 \otimes \cS[ \cD^{(2)}_{J}]
\ ,
\eeq
where $ \cD^{(2)}_{J}$ is the discontinuity of the two-loop remainder in the $P_{J}^2$-channel,  and the sum is extended to all channels where there is a physical discontinuity.
\item[{\bf 3.}] The second and the last entries of the symbol must also satisfy certain additional constraints, to be discussed below.
\item[{\bf 4.}]
The symbol must have trivial   collinear limits.
\item[{\bf 5.}]
The symbol must satisfy all relevant symmetries.
\item[{\bf 6.}]
The ansatz must be the symbol of a local  function, which in turn means that the symbol must satisfy certain integrability conditions.
\end{itemize}
We now focus on the three-point case and discuss the implementation of the properties listed above.

{\bf 1.} To begin with,
we wish to claim that the symbol of the three-point form factor remainder function takes its entries only from the following set of six kinematic variables
\beq
\label{only}
\{u,v,w; 1-u, 1-v, 1-w\}
\ ,
\eeq
where $u$, $v$ and $w$ are the scale invariant ratios defined in \eqref{uvw}.
This statement is highly non-trivial and is an important constraint, as it does not allow for dependence on any other variables such as $1+u$, or $1+uv$ etc. Our claim is substantiated by explicit inspection of the  two-loop master integrals
with one  off-shell and three on-shell legs  \cite{grp,grnp}  in terms of which our two-loop, three-point  form factors is expanded. We have calculated the symbol of the relevant integral functions, with the result that they always depend on the variables listed in \eqref{only} only and that the first entry of the symbol is always taken from the subset $\{u, v, w \}$.

{\bf 2.} The first-entry condition requires that the first entry of the symbol must  be either $u$, $v$, or $w$, but the remaining three variables $1-u$, $1-v$, $1-w$ cannot appear as possible first entries.
This is also confirmed by inspection of the integral functions appearing
in \cite{grp,grnp} as mentioned under {\bf 1.}

{\bf 3.} Furthermore, the second entry of the symbol is restricted in the following way \cite{pulling}.
If the first entry is $u$, then the second entry can be $1-u$ or $v$ or $w$, but not $u$ or $1-v$ or $1-w$; similar conditions in the case when first entry is $v$ or $w$ are obtained by cyclic permutations of $u$, $v$ and $w$.
Finally, we impose a last entry condition on the symbol  \cite{Dixon:2011nj,CaronHuot:2011ky}, by requiring it to be always  of the form  $u/(1-u)$, or $v/(1-v)$, or $w/(1-w)$.
If we insist in using the variables in \eqref{only} as entries for our symbol, the above conditions relate pairs of coefficients, namely the coefficient of a term in the symbol whose last entry is $u$ is  the opposite of  the coefficient of the term whose first three entries are the same, with  the last one equal to  $1-u$. And similarly for the pairs $v$ and $1-v$, and $w$ and $1-w$.

At this point, we observe that imposing the first three conditions, one obtains an ansatz for the symbol in terms of 324 independent parameters.

{\bf 4.} Next, we recall that the three-point remainder must vanish in all collinear limits.%
\footnote{We would like to thank Paul Heslop for a very useful conversation on collinear limits of symbols.}
Imposing this constraint term by term on the symbol is too restrictive, as there can be cancellations among different terms. Therefore, we keep the most general ansatz and impose that  it vanishes in any of the three simple collinear limits. We also recall that,  in order for a symbol to vanish,  it is enough that one of its entries smoothly goes to one in the limit.

For definiteness, let us focus on the $s_{12}\to 0$, or $u\to 0$ limit, which we take as follows:  we leave untouched every entry equal to $u$ in the symbol; however we can replace $1-u \to 1$, and furthermore, since $v+w=1$ in the collinear limit, we can replace $w\to 1-v$. Requiring the symbol to vanish in this limit, we obtain three sets of equations,  for the three independent simple collinear limits $s_{12}\to 0$ (or $u \to 0$), $s_{23}\to 0$ (or $v\to 0$) and $s_{31}\to 0$ ($w\to 0$).  Altogether, the collinear equations impose 99 independent constraints on the symbol.

{\bf 5.} Next, we impose that the remainder $\cR_3^{(2)}$ must be totally symmetric in $u$, $v$, $w$. This leads to 52 linear relations between the coefficients.

 {\bf 6.} Finally, we impose the integrability condition on the ansatz, namely the constraint that the symbol is derived from a local function. This is implemented as follows \cite{gonch,pulling}: one takes the symbol
 \beq
 \cS \ = \ \sum w_1 \otimes \cdots \otimes w_n
 \ ,
 \eeq
and replaces two consecutive entries  $w_i$,  $w_{i+1}$ by $dw_i\wedge dw_{i+1}$. The resulting expression must vanish:
  \beq
 \sum   dw_i\wedge dw_{i+1} \Big(w_1 \otimes \cdots \otimes w_{i-1} \otimes w_{i+2}  \otimes \cdots \otimes w_n \Big)\ = \ 0
 \ .
 \eeq
  There are three different integrability conditions we have to impose in our case.%
  \footnote{In general, there are  $p-1$ integrability conditions for a transcendentality $p$ symbol.}
  Remarkably, by requiring these  to be obeyed, we find that there is a unique solution for the symbol of the two-loop remainder (up to an overall normalisation). It is given by the following  compact expression:
\beqa
\label{nice}
\cS^{(2)} & = &
-2 u\otimes (1-u)\otimes (1-u)\otimes \frac{1-u}{u}
+u\otimes (1-u)\otimes u\otimes \frac{1-u}{u}
\nonumber \\
&&
   -u\otimes (1-u)\otimes v\otimes \frac{1-v}{v}
   -u\otimes (1-u)\otimes w\otimes \frac{1-w}{w}
      \nonumber \\
   &&
   -u \otimes v \otimes (1-u)\otimes \frac{1-v}{v}
   -u \otimes v \otimes (1-v)\otimes \frac{1-u}{u}
   \nonumber \\
   &&
   +u\otimes v \otimes w \otimes \frac{1-u}{u}
   +u\otimes v \otimes w \otimes \frac{1-v}{v}
   \nonumber \\
   &&
      +u\otimes v\otimes w\otimes \frac{1-w}{w}
   -u\otimes w\otimes (1-u)\otimes\frac{1-w}{w}
\nonumber \\
&&
   +u\otimes w\otimes v\otimes \frac{1-u}{u}
   +u\otimes w\otimes v\otimes \frac{1-v}{v}
\nonumber \\
&&
   +u\otimes w\otimes v\otimes \frac{1-w}{w}
   -u\otimes w\otimes (1-w)\otimes
   \frac{1-u}{u}
   \nonumber \\
&& + \ \mathrm{cyclic} \ \mathrm{permutations} \, .
\eeqa

The next challenge is twofold: firstly, we wish to determine   the function whose symbol is given by \eqref{nice}; and secondly, we wish to determine terms missed by the symbol, e.g.~terms of the form $\pi^2 \times F_2$ where $F_2$ is a sum of transcendentality-two functions with rational coefficients.

In this respect, there is an additional piece of information about \eqref{nice} that we would like to mention. Our symbol  $\cS^{(2)}$  defined in \eqref{nice} satisfies an important symmetry constraint \cite{gonch} discussed in \cite{symbol},  namely
\beq\label{symsymm}
\cS^{(2)}_{abcd} - \cS^{(2)}_{bacd} - \cS^{(2)}_{abdc}+ \cS^{(2)}_{badc} - (a \leftrightarrow c\, ,\,  b \leftrightarrow d)\ = \ 0
\ .
\eeq
According to a conjecture of Goncharov,  symbols with this peculiar property can always be
obtained from a function involving logarithms
and classical polylogarithms $\mathrm{Li}_k$'s with $k\leq 4$ only \cite{gonch, symbol}.
The explicit solution we will present in the next section will confirm this expectation
beautifully.
As we will show in the final part of this paper, there is an alternative way to obtain an analytic result
of the form factor remainder in terms of two-dimensional harmonic polylogarithms  \cite{Remiddi:1999ew}. This is due to a remarkable relation between the $\cN=4$ form factor and the planar, maximally transcendental part of the two-loop QCD amplitude for $H \to ggg$ recently obtained in \cite{Nigel, kkt}.

\subsection{The analytic remainder function}
The remaining task now  is  to find a transcendentality-four function whose symbol is given by \eqref{nice}. Recall that the symbol only takes entries from the list $\{u,v,w,1-u,1-v,1-w\}$ and
has the symmetry \eqref{symsymm},  which implies the result should be expressed purely in terms of classical polylogarithms of degree up to four and logarithms \cite{gonch,symbol}. This however does not fix a priori the allowed arguments of these functions, but the arguments of individual functions must be such that the symbol of that function has only entries from that list. Taking these considerations into account, the most general ansatz will be built from the following set of
functions:
\be
\log x_1 \log x_2 \log x_3 \log x_4 \ , \ \mathrm{Li}_2(x_1) \log x_2 \log x_3 \ , \ \mathrm{Li}_2(x_1) \mathrm{Li}_2(x_2)
\ , \ \mathrm{Li}_3(x_1) \log x_2 \ \mathrm{and} \ \mathrm{Li}_4(x_i) \ ,
\ee
where we found it sufficient to take the possible arguments $x_i$ from the list
\be\label{args}
\left\{
u,v,w,1-u,1-v,1-w,1-\frac{1}{u},1-\frac{1}{v},1-\frac{1}{w}, -\frac{u v}{w}, -\frac{v w}{u},-\frac{w u}{v}
\right\} \ .
\ee
Imposing the constraint that the ansatz has the same symbol as \eqref{nice} one can easily
find a solution. We have then applied various polylogarithm identities to simplify the {\it raw} solution obtained in this way. The final result takes the remarkably simple and compact form
\beqa\label{beauty}
\mathcal{R}^{(2)}_3 & = &  -2 \left[ \mathrm{J}_4 \left( -\frac{u v}{w}\right)+\mathrm{J}_4 \left( -\frac{v w}{u}\right)+\mathrm{J}_4 \left( -\frac{w u}{v}\right)\right] -8 \sum_{i=1}^3 \left[ \mathrm{Li}_4 \left(1-u_i^{-1}\right)+\frac{\log^4 u_i}{4!} \right] \nonumber\\
&& 
-2 \left[ \sum_{i=1}^3 \mathrm{Li}_2 (1-u_i^{-1}) \right]^2
+\frac{1}{2} \left[ \sum_{i=1}^3 \log^2 u_i\right]^2  -\frac{\log^4(u v w)}{4!} - \frac{23}{2} \zeta_4   \  , \nonumber \\
&&
\eeqa
where $u_1=u$, $u_2=v$ and $u_3=w$ and we have introduced the function
\be
\mathrm{J}_4(z) := \mathrm{Li}_4(z)-\log(-z) \mathrm{Li}_3(z)+\frac{\log^2(-z)}{2!} \mathrm{Li}_2(z)-\frac{\log^3(-z)}{3!} \mathrm{Li}_1(z) - \frac{\log^4(-z)}{48} \ .
\ee
It is curious to note here that $\mathrm{J}_4(z)$ is almost identical to the function $\mathrm{D}_4(z)$  introduced by Ramakrishnan. The functions $\mathrm{D}_m(z)$, $m > 2$,
are generalisations of the Bloch-Wigner functions (see \cite{zagier} for an inspirational exposition of these topics and references).

In the representation obtained above we have already taken into account beyond-the-symbol ambiguities which arise due to the fact that the symbol is blind to transcendentality-four terms of the form $\pi^4$ or $\pi^2 \times \{ \log x_i \log x_j, \mathrm{Li}_2(x_i)\}$. It is a simple task to fix these ambiguities using constraints from permutation symmetry and collinear limits. In our case it was sufficient to add the $\zeta_4$ term to get a symmetric function, that is smooth throughout the Euclidean region defined as $0 \leq u \leq 1, 0\leq v  \leq 1, 0\leq w \leq 1$ and $u+v+w=1$, and vanishes in all collinear and soft limits.

Finally, we have collected in Table 2 results from our numerical evaluations in Section \ref{3.1} and compared them with the
exact result \eqref{beauty}. This also serves as confirmation of the overall normalisation of
the remainder,  which is not fixed by the symbol alone.

\begin{table}[h]
\begin{center}
\begin{tabular}{|c|c|c|c|}
\hline
$(u,v,w)$ & numerical $\cR_3^{(2)}$ & est. error & analytic $\cR_3^{(2)}$\\
\hline \hline
(1/3, 1/3, 1/3) & -0.1519   & 0.02 & -0.148966\\
(1/4, 1/4, 1/2) & -0.1203   & 0.02 & -0.134873\\
(1/5, 2/5, 2/5) & -0.1301   & 0.02 & -0.136454\\
(1/2, 1/3, 1/6) & -0.1080   & 0.03 & -0.125366\\
\hline
\end{tabular}
\caption{\it Comparison of numerical and analytic values of the remainder $\cR_3^{(2)}$.}
\end{center}
\end{table}

\subsection{Connections  between amplitude and form factor remainders}
We have seen earlier that the three-point form factor remainder depends on three variables $u$, $v$ and $w$ defined in \eqref{uvw},  where $u+v+w=1$. In \eqref{nice} we have presented the symbol of this remainder.  
An a priori entirely different calculation is that of the six-point MHV amplitude remainder,  whose  symbol  was explicitly evaluated in \cite{symbol} using the results of \cite{vittorioetal}.  This symbol can be expressed in terms of the three independent cross-ratios $u$, $v$ and $w$ one can write down with six lightlike momenta satisfying $\sum_{i=1}^6 p_i =0$, as well as three additional variables $y_u$, $y_v$, $y_w$. An explicit compact expression for this six-point amplitude remainder symbol is given in Eq.~(21) of  \cite{Dixon:2011pw}.%
\footnote{The  definitions of the $y$ variables can also be found in Eq.~(15) of \cite{Dixon:2011pw}. We will not need their explicit expressions here.}
It can be seen from that formula that the symbol is naturally decomposed into  two terms, 
\beq
\label{conley}
\mathcal{S}_{6, \, {\rm ampl}}^{(2)} \ = \ \hat{\mathcal{S}}_{6,\, {\rm ampl}}^{(2)}(u, v, w) \, + \, \tilde{\mathcal{S}}_{6,\, {\rm ampl}}^{(2)}(u, v, w; y_u, y_v, y_w)
\ .
\eeq
The second term in \eqref{conley} is given by \cite{Dixon:2011pw}
\beqa
\tilde{\mathcal{S}}_{6,\, {\rm ampl}}^{(2)}(u, v, w; y_u, y_v, y_w) &=&  - {1\over 8} \big[ 
 u\otimes (1-u) \otimes y_u y_v y_w \, - \, 2 u \otimes v\otimes y_w\big] \otimes y_u y_v y_w \nonumber  \\ 
 & + & {\rm permutations.}
 \eeqa
On the other hand, it is straightforward to check that $\hat{\mathcal{S}}_{6,\, {\rm ampl}}^{(2)}(u, v, w)$ is in fact identical (up to an overall multiplicative constant)  to the symbol of our two-loop three-point form factor remainder \eqref{nice}  upon identifying the three (unconstrained) cross-ratios of the amplitude calculation with our three (constrained) variables $u$, $v$ and $w$ defined in \eqref{uvw}.  

Note that   
$\hat{\mathcal{S}}_{6,\, {\rm ampl}}^{(2)}(u, v, w)$ has trivial collinear limits both when it is part of a  form factor symbol, where  $u+v+w=1$, and when it is part of the six-point remainder symbol. In the latter case   the three cross-ratios $u$, $v$ and $w$ are in general unconstrained, but in the collinear limit they become dependent -- for instance, in the collinear limit  $w\to 0$ one also has that $u+v \to 1$.

This  is reminiscent of a similar interesting coincidence observed at strong coupling in \cite{mz}. There, the four-point form factor remainder was evaluated in  the particular case of $(1+1)$-dimensional kinematics, and found to be expressible in terms of the eight-point MHV amplitude remainder in $(1+1)$-kinematics, which was  determined in \cite{amsmalloctopus,amoctopus}. 
We note that the relations between form factor and amplitude remainders at strong and weak coupling seen in  \cite{mz} and in the present paper can by no means be anticipated from the explicit calculation. For instance, in our weak coupling calculation we find that the three-point form factor remainder  contains several integral functions -- in particular non-planar ones -- which do not appear in the corresponding six-point amplitude calculation. 

Incidentally, we  note  that while the four-point form factor remainder and the eight-point MHV amplitude remainder in of $(1+1)$-dimensional kinematics depend both on two independent variables,%
\footnote{Recall that for this restricted  kinematics, the $2n$-point form factor remainder depends on $2n - 2$ variables, whereas the amplitude remainder depends on $2n-6$ cross-ratios.}
in our case the number of independent variables is different -- the form factor remainder depends on two independent simple ratios, and the amplitude's on three cross-ratios.

Perhaps this connection between amplitude and form factor remainders is another indication that the paradigm ``loop observable = rational coefficient $\times $ integral function" may obscure the  simplicity of the final result of a loop calculation. 
Furthermore it is very likely that the symbol technology will allow to simplify the expressions of the  (higher-loop) integral functions themselves, such as those  appearing in   \eqref{nice} and in QCD. For a recent example in QCD see \cite{mant}.

\subsection{A surprising relation with QCD}
In this final section we wish to discuss an intriguing connection  of our result with the recent
work of \cite{Nigel}. In that paper, the two-loop helicity amplitudes for $H \to ggg$ and $H \to q\bar{q} g$ were computed in the large top mass limit. In this approximation the top quark can be integrated out at one loop and produces a new effective vertex of the form $H g g$. If we consider the process $H \to ggg$, which is the case of interest for our discussion here, the calculation is equivalent to that of a three-gluon form-factor with a $\Tr F^2$ operator insertion,  where $F_{\mu\nu}$ is the gluon field strength. As we noted earlier, the operator $\Tr \phi_{12}^2$ --  whose form factor we consider in this paper  -- is the lowest component of the stress-tensor multiplet, which among many other operators contains the on-shell Lagrangian $\mathcal{L}_{\mathrm{o.s.}}=\Tr F_{\mathrm{SD}}^2 +\cdots$,
where $F_{\mathrm{SD}}$ is the self-dual part of the field strength,  and the dots stand for cubic and quartic terms in the Lagrangian. Up to a universal helicity-dependent prefactor the form factor of the operator with three-gluons $\langle g^-(p_1) g^-(p_2) g^+(p_3) |\mathcal{L}_{\mathrm{o.s.}}(0)|0\rangle$ is equal to the form factor we have been discussing  in this paper, and on the other hand is related to the object considered in \cite{Nigel} in QCD.

The full QCD result of \cite{Nigel} is of course very complicated and has no resemblance to
\eqref{beauty}. The story becomes more interesting if we focus on the planar part of the finite
two-loop remainder $A^{(2)}_\Omega$ defined in  (5.8) and (5.20) of \cite{Nigel},  and with explicit formulae in terms of 2dHPL's given in Eqns.~(B.1) and (B.7) of the same reference for the two different gluon helicity configurations.%
\footnote{Note that $A^{(2)}_\Omega$ are the coefficients of the $N^2$ terms in the finite remainder defined in (5.20) of \cite{Nigel}. The QCD result with external gluons does not have subleading in $N$ terms, but does have contributions from quarks which are proportional to powers of the number of quark flavours $N_F$. In order to compare to the $\cN=4$ form factor calculation we have suppressed such $N_F$-dependent terms. 
Diagrams in the QCD calculation where gluons run in the loops are of course in common with the $\cN=4$ form factor calculation, and are accounted for by $A^{(2)}_\Omega$.  }
It is important to note that these finite remainders are defined using the formalism of \cite{magician} widely used in QCD to remove the universal two-loop infrared divergences,  and are different from the BDS finite remainders used in this paper \cite{abdk,bds}; in particular the former does not vanish in collinear limits. However, it is easy to relate the two definitions in $\cN=4$ (see \cite{abdk}), because one simply has to subtract
$\frac{1}{2}(F^{(1)}_{\mathrm{fin}})^2 -2 \zeta_2 F^{(1)}_{\mathrm{fin}}$ from Catani's remainder \cite{magician} in order to arrive at the BDS remainder.

The interesting observation arises as follows:
First, take the finite two-loop remainders
$A^{(2)}_\alpha$ or $A^{(2)}_\beta$ from Eqns.~(B.1) and (B.7) of \cite{Nigel}, corresponding to the helicity configurations $(g^+g^+g^+)$ and $(g^+g^+g^-)$ for the external gluons, respectively, and extract the terms containing functions of maximal transcendentality degree (four in this case). Then map them to BDS remainders using the prescription outlined above with $F^{(1)}_{\mathrm{fin}}$ replaced by the degree-two parts of the one-loop amplitudes $A^{(1)}_\alpha$ and $A^{(1)}_\beta$ (Eqns.~(A.1) and (A.4) in \cite{Nigel}), which coincides up to a factor of two with the $\eps^0$ term of our one-loop form factor in \eqref{3pt1loop}.
Finally, calculate the symbol of the resulting expression.  Surprisingly,  one finds exactly \eqref{nice},  for both helicity configurations.%
\footnote{In this respect we note that in $\cN=4$ SYM the two form factors with MHV $(g^+g^+g^-)$ and maximally non-$\overline{\mathrm{MHV}}$ $(g^+g^+g^+)$ helicity configurations are actually related, a fact that was explained in Section 3.2.3 of \cite{harmony}. }
This implies that our form factor remainder \eqref{beauty} is contained  in the full QCD result of \cite{Nigel},  and allows to present the latter  in a more compact form. It would be interesting to see if symbols can be used effectively to simplify the terms of lower degree (transcendentality) in the expressions of \cite{Nigel},  or in other QCD results.

Alternatively, this connection provides us with  an additional strategy to reconstruct a function from the symbol \eqref{beauty}  by simply projecting out the functions of maximal degree and performing the subtraction described above to get the BDS remainder. The result
obtained in this way is given in terms of 2dHPL's and the remaining beyond-the-symbol ambiguities can be fixed by considering symmetries and collinear limits. In this case the necessary correction terms can also be read off directly from e.g.~$A^{(2)}_\beta$ in (B.6) of \cite{Nigel}: one has to keep all terms of the form $\pi^2 \times F_2$  where $F_2$ is a sum of degree-two functions, while terms of the form $\zeta_3 \times F_1$ where $F_1$ is a sum of logarithms have to be dropped because they spoil the correct collinear behaviour. We also note that these correction terms are essential to restore  the symmetry of the remainder $\mathcal{R}^{(2)}_3$ under permutations of $(u, v, w)$.   We have checked that the two formulae for the remainder match exactly up to a factor of four.%
\footnote{The relative factors of four and two for two-loop  and one-loop quantities respectively are due to different normalisation conventions of the gauge coupling constant.}

We believe that  this is the first example where the principle of maximal transcendentality,
first observed in \cite{maxtrans} for anomalous dimensions of operators, relates physical quantities with non-trivial kinematic dependence in planar QCD and $\cN=4$ SYM. It would be exciting to identify more such examples and we are confident that the technology of symbols will turn into a powerful tool also in QCD.


\section*{Acknowledgements}

We would like to thank Zvi Bern, Rutger Boels, Paul Heslop, Valya
Khoze, Lorenzo Magnea and Bill Spence  for illuminating discussions. 
This work was supported by the STFC Grant ST/J000469/1,  
``String theory, gauge
theory \& duality", and by the German Science Foundation (DFG)
within the Collaborative Research Center 676 ``Particles, Strings
and the Early Universe". AB thanks the ``Feinberg Foundation
Visiting Faculty Program" at the Weizmann Institute of Science and
Tel Aviv University for hospitality. GT acknowledges the warm
hospitality and support from the Institute for Particle Physics
Phenomenology, Durham University,  through an IPPP Associateship, as
well as from the Weizmann Institute of Science.

\appendix

\section{Complete one-loop amplitudes with fundamental and adjoint generators}
At one loop, complete (planar plus non-planar) amplitudes can be written as \cite{bddk}
\beqa
\label{cd}
\mathcal{A}^{(1)}(1, \ldots , n ) &= & \mathcal{A}^{(1)}_{\rm P}(1, \ldots , n ) \, + \, \mathcal{A}^{(1)}_{\rm NP} (1, \ldots , n )
\nonumber \\
&=&
N\, \sum_{\sigma\in S_n / \mathbb{Z}_n}\!\!\Tr (T^{a_{\sigma_1}} \cdots T^{a_{\sigma_n}} ) \, A_{n; 1}^{(1)} (\sigma_1, \ldots, \sigma_n) \\
&+&
  \sum_{\sigma\in S_n / S_{n; c} }\!\!\sum_{c=2}^{\lfloor n/2\rfloor + 1}\!\!\Tr (T^{a_{\sigma_1}} \cdots T^{a_{\sigma_{c-1}}} )\Tr (T^{a_{\sigma_c}} \cdots T^{a_{\sigma_n}} ) \,  A_{n; c}^{(1)} (\sigma_1, \ldots, \sigma_n)\, .
\nonumber
\eeqa
Here $S_n$ is the set of permutations of $n$ objects, and $S_{n; c}$ is the subset of permutations which leaves the double-trace structure in \eqref{cd} invariant. $A_{n; 1}^{(1)} (\sigma_1, \ldots, \sigma_n)$ are colour-ordered one-loop amplitudes, whereas $A_{n; c}^{(1)} (\sigma_1, \ldots, \sigma_n)$ are certain linear combinations thereof, constructed as \cite{bddk}
\beq
\label{ccc}
A_{n;c}^{(1)} (1,2,\ldots ,c-1; c,c+1, \ldots ,n) = (-1)^{c-1}  \sum_{\sigma\in \mathrm{COP}\{\alpha\}\{\beta\}} A_{n;1}^{(1)}(\sigma) \ ,
\eeq
where $\alpha_i \in \{\alpha\} \equiv \{c-1,c-2,\ldots,2,1\}$, $\beta_i \in \{\beta\} \equiv \{c,c+1,\ldots,n-1,n\}$, and $\mathrm{COP}\{\alpha\}\{\beta\}$ denotes the set of all permutations of $\{1,2,\ldots,n\}$ where $n$ is held fixed, and such that that the cyclic ordering of the $\alpha_i$ within $\{\alpha\}$ and of the $\{\beta_i\}$ within $\{\beta\}$ is maintained, while allowing for all possible relative orderings of the $\alpha_i$ with respect to the $\beta_i$.

One can also work with the alternative, remarkably compact  representation of the complete one-loop amplitudes presented in \cite{dddm},
\beq
\label{vitto}
\mathcal{A}^{(1)}(1, \ldots , n)\ = \  \sum_{\sigma\in S_n / (\mathbb{Z}_n \times \mathcal{R})} {\rm Tr} (F^{a_{\sigma_1}}\cdots F^{a_{\sigma_n}})\, A_{n; 1}^{(1)}  (\sigma_1, \ldots , \sigma_n) \, ,
\eeq
where $F^a_{bc}$ are adjoint generators.%
\footnote{We define the adjoint generators are defined as $F^a_{bc} := -i f^{abc}$, where the structure constants $f^{abc}$ have an additional factor of $\sqrt{2}$ compared to the usual ones. This normalisation is chosen in order to accommodate for the normalisation
$\Tr (T^a T^b) = \delta^{ab}$ of the fundamental generators. }
The sum in \eqref{vitto} contains $(n-1)! / 2$ terms.  Note that planar and non-planar contributions are both contained in  \eqref{vitto}, where only colour-ordered amplitudes are summed.



\end{document}